\documentclass[9.5pt,journal,compsoc]{IEEEtran}
\ifCLASSOPTIONcompsoc
\usepackage{url}
  \usepackage[nocompress]{cite}
\else
  \usepackage{cite}
\fi
\usepackage{booktabs}
\usepackage{multirow}
\ifCLASSINFOpdf
   \usepackage[pdftex]{graphicx}
   \DeclareGraphicsExtensions{.pdf,.jpeg,.png}
\else

   \usepackage[dvips]{graphicx}
   \graphicspath{{../images/}}
   \DeclareGraphicsExtensions{.png}
\fi

\usepackage{xcolor}
\usepackage{todonotes}
\usepackage{makecell}
\usepackage{multirow}
\usepackage[normalem]{ulem}

\useunder{\uline}{\ul}{}

\begin{document}

\title{Beyond One-Size-Fits-All: A Survey of Personalized Affective Computing in Human-Agent Interaction}

\author{Jialin Li, Maha Elgarf, Alia Waleed Mahmoud, Hanan Salam

\IEEEcompsocitemizethanks{\IEEEcompsocthanksitem H. Salam is with the Social Machines \& Robotics (SMART) Lab, and the Center of AI \& Robotics (CAIR), New York University, Abu Dhabi, E-mail: hanan.salam@nyu.edu
\IEEEcompsocthanksitem J.Li is with SMART Lab, New York University, Abu Dhabi, E-mail: jl10897@nyu.edu
\IEEEcompsocthanksitem M. Elgarf is with SMART Lab, New York University, Abu Dhabi, E-mail: mae9866@nyu.edu
\IEEEcompsocthanksitem A. Waleed is with SMART Lab, New York University, Abu Dhabi, E-mail: awm330@nyu.edu}}


\maketitle

\begin{abstract}
In personalized machine learning, the aim of personalization is to train a model that caters to a specific individual or group of individuals by optimizing one or more performance metrics and adhering to specific constraints. In this paper, we discuss the need for personalization in affective computing and present the first survey of existing approaches for personalization in affective computing. Our review spans training techniques and objectives towards the personalization of affective computing models across various interaction modes and contexts. We develop a taxonomy that clusters existing approaches into Data-level and Model-level approaches. Across the Data-Level and Model-Level broad categories, we group existing approaches into seven sub-categories: (1) User-Specific Models, (2) Group-Specific Models, (3) Weighting-Based Approaches, (4) Feature Augmentation, (5) Generative-Based Models which fall into the Data-Level approaches, (6) Fine-Tuning Approaches, and (7) Multitask Learning Approaches falling under the model-level approaches. We provide a problem formulation for personalized affective computing, and to each of the identified sub-categories. Additionally, we provide a statistical analysis of the surveyed literature, analyzing the prevalence of different affective computing tasks, interaction modes (i.e. Human-Computer Interaction (HCI), Human-Human interaction (HHI), Human-Robot Interaction (HRI)), interaction contexts (e.g. educative, social, gaming, etc.), and the level of personalization among the surveyed works. Based on our analysis, we provide a road-map for researchers interested in exploring this direction.
 
\end{abstract}

\section{Introduction}
 Affective computing is a sub-field of Human-Agent Interaction (HAI) that aims at developing computational models to recognize and interpret human affective states (e.g. emotions, engagement, pain, etc.). 
Examples of affective computing tasks include emotion recognition \cite{Rescigno202035811,mukherjee2021intense}, facial expressions recognition \cite{hazourli2021multi,soladie2012multimodal}, Action Units (AU) detection \cite{senechal2012facial,senechal2011combining}, engagement detection \cite{salam2023automatic,salam2015engagement}, pain detection \cite{lopez2018multi}, and mental health related affective tasks such as depression recognition \cite{muzammel2021end,Muzammel2020}, or anxiety detection \cite{bendevski2021towards}, among others. 

Affective computing approaches typically focus on recognizing and interpreting affective states expressed through verbal and non-verbal signals (e.g. body movements, facial expressions, or speech) \cite{wang2022systematic,karpouzis2007modeling}   or through the detection of physiological changes in the body that are associated with different affective states \cite{lee2021current}. It involves using machine learning algorithms to analyze patterns and predict such states. 
Existing approaches use deep learning on raw data \cite{gupta2023analysis,pang2015deep,jiang2014predicting}, or traditional machine learning approaches on hand-crafted features \cite{arifin2008affective,soleymani2009bayesian,teixeira2012determination}. 

Personalized machine learning refers to the use of machine learning algorithms to create personalized computational models that are customized to the characteristics of an individual user or a group of users that share common characteristics \cite{mcauley2022,schneider2021personalization}. This approach can help to improve the accuracy and effectiveness of machine learning systems by tailoring them to the specific characteristics of each individual user or group of users. Existing personalization techniques are mostly in the areas of web mining, recommender systems, fashion, dialog, and personalized health \cite{mcauley2022}. 

The importance of personalized models for affective computing is very well supported by evidence in social and behavioral sciences and HAI research\cite{reichardt2020affective}. 
For instance, prior studies have shown that there are variations in personality characteristics across various user profiles, including age ranges, genders, and even cultural backgrounds~\cite{akyunus2021age,weisberg2011gender,hang2021age}, affecting how individuals show or manifest their emotional states. 

Hence, personalization of affective computing models entails taking into account individual differences when developing such models, resulting in interactions that resonate more effectively with users. 
By adapting to individual nuances, socially intelligent agents become more adept as companions to the users in various domains, from healthcare to education.

Personalizing computational models of affect can also enhance personalization strategies in HAI, for example user interface design in HCI, which aims to improve user experience (e.g. ease of finding information, satisfaction, long-term engagement), or robot's gender and personality in HRI. 
 
Despite the aforementioned evidence of the importance of personalized computational models, the HAI community has primarily concentrated on universal approaches for the computational models of affect, overlooking personalized models \cite{Kathan2022}. While there have been a few approaches attempting to develop customized models for affective computing tasks related to recognizing emotions \cite{vogt-andre-2006-improving,yusuf2017individuality,chu2016selective}  and predicting personality traits \cite{shao2021personality}, there is no comprehensive review of the field. 

Thus, there is a significant gap in the understanding of personalized affective computing models and their potential applications. A recent book \cite{mcauley2022} explored approaches for personalized machine learning, focusing  however on personalization in the context of recommender systems. 

A literature review could provide a deeper understanding of the existing research in personalized affective computing, identify the gaps and limitations of current techniques, and suggest future directions for study in this field.

In this paper, we conduct a thorough review of current literature on personalized affective computing. We present a new taxonomy of the existing methods, mathematical formulations and graphical representations of each method. We also conduct a statistical analysis of the surveyed literature, from a set of five angles: (1) the personalization technique used (user-specific modeling, fine-tuning approach, feature augmentation...etc), (2) the stage where the personalization is performed (data-level vs. model-level), (3) the target that the personalization algorithm is designed for (user vs. group), (4) the interaction mode (HCI, HHI, HRI), and (5) the personalization context (emotion understanding, health, education...etc), serving as a guide for future studies in developing personalized affective systems. 

To the best of our knowledge, this paper is the first comprehensive survey focusing on personalization techniques of affective computing. We believe that our work will contribute to an improved environment for future research in this field.

In summary, the contributions of this paper are the following: 
\begin{itemize}
\item First comprehensive survey of personalization techniques for affective computing.
\item New taxonomy for personalized affective computing. 
\item Problem formulation of personalized affective computing, as well as the different identified categories.
\item Analysis of existing personalization techniques across different contexts and modes of interaction, leading to the identification of existing gaps in the literature and informing future research directions.
\end{itemize}

The paper is organized as follows: Section \ref{sec:method} describes the method and inclusion criteria. 
Section \ref{sec:personalfactors} discusses the need for personalization approaches in affective computing. 
Section \ref{sec:problem-formulation} presents a mathematical formulation of the personalized affective computing problematic, 
and explains our taxonomy and categorization of the personalized affective computing techniques reviewed.
Section \ref{sec:discussion_openQ} discusses and presents a statistical analysis of the surveyed papers and addresses open questions for further research.
Section \ref{sec:conclusion} concludes the paper.

\section{Method \& Inclusion Criteria}
\label{sec:method}

In this review, we aimed to survey the literature in the field of personalised affective computing by attempting to answer the following research questions:
\\\textbf{ (1) How can we categorize the different techniques used in the surveyed literature of personalized affective computing?}
\\To answer this question, we present a taxonomy of the deep learning and machine learning techniques used in the field under seven different categories in section \ref{sec:problem-formulation}. Furthermore, we clarify the techniques by providing a mathematical forumaltion and a graphical representation for each technique with some examples from the literature.
\\\textbf{ (2) What are the typical trends observed in the surveyed literature of personalized affective computing?}
\\We aimed to answer this question by conducting a statistical analysis of the reviewed papers through a set of five angles: personalization techniques, personalization stage, personalization level, interaction mode, and personalization context. Moreover, we present a thorough explanation of the trends in the use of machine learning versus deep learning techniques and data-level versus model-level techniques in the reviewed literature. All details are provided in section \ref{sec:discussion_openQ}. 


Our initial search was conducted in Google Scholar using the following terms: (``personalized'' OR ``individualized'') AND (``affective computing'' OR ``emotion'' OR ``engagement'' OR ``facial expression'' OR ``sentiment''). Additional relevant papers were identified from the reference lists of papers retrieved through the initial search. Papers were screened for inclusion based on the following predefined criteria: 
\begin{itemize}
\item Tackles tasks that are affective in nature or have a major affective component. For instance, engagement is a construct that has affective, cognitive, and behavioral components, with some works treating it as a purely affective construct in some contexts \cite{salam2023automatic}. Similarly, personality recognition is a task that involves an affective component \cite{vinciarelli2014survey}. Consequently, papers revolving on personalized computing approaches for these tasks were included in this paper. 
\item Tackles algorithmic computation of affective tasks. Papers that tackle personalization or adaptation of interface (e.g. robot, agent, etc.) are excluded. 
    \item Is a full-text research article: reviews, tutorials, abstracts, posters, and other non-research paper formats were excluded.
    \item Reports on personalization techniques targeting affective goals: only the personalization techniques described to directly serve affective tasks were included.
\end{itemize}
The initial search identified 124 candidate papers published between 2006-2024. After screening based on the inclusion and exclusion criteria, 23 papers were excluded, resulting in 101 papers to be included in the literature review.

The included papers were further analyzed by the authors through weekly discussions to develop the taxonomy categories presented in this paper. The primary author employed a shared spreadsheet to document paper annotations and perform statistical analysis, enhancing the efficiency of the process and ensuring it remains up-to-date. The aim of the statistical analysis is to describe the state of the literature that was surveyed (e.g., the distribution of tasks in the cited literature).

All 101 included papers contributed to the overall discussion and statistics in this survey. However, due to space limitations, only a representative subset of these papers are cited in detail where relevant concepts or techniques are discussed. The full reference list includes all 101 papers that were reviewed.

\section{The Need for Personalization in Affective Computing: Personal Factors}
\label{sec:personalfactors}

\begin{figure}
    \centering
    \includegraphics[width=0.7\linewidth]{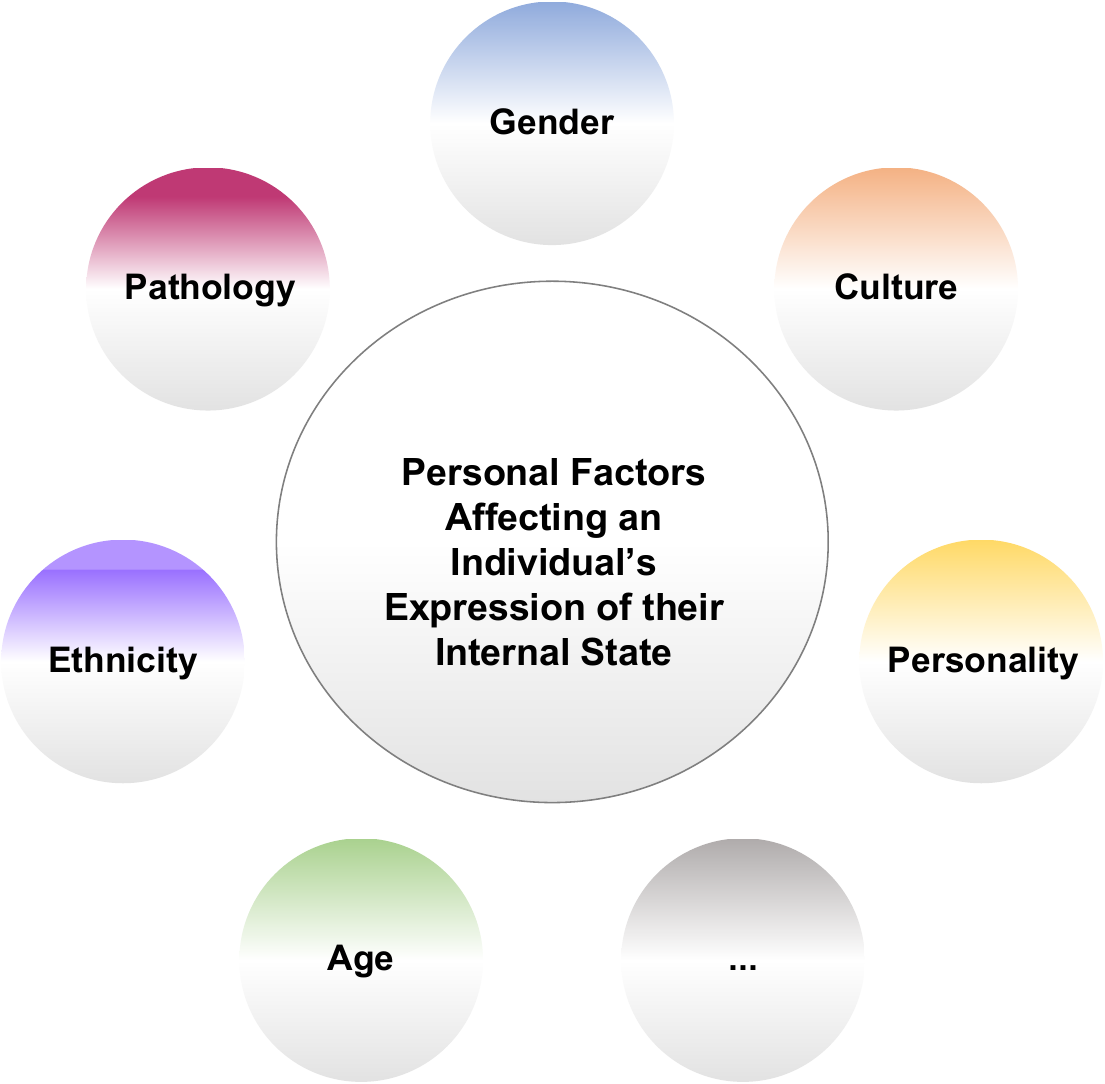}
    \caption{Example of some personal factors that may have an effect on an individual's expression of their affective state.}
    \label{fig:personal-factors}
\end{figure}

In this section, we highlight some personal (subjective) factors that were shown to have an effect on an individual's expression of their affective states. These factors refer to the characteristics and traits that are unique to each individual, or group of individuals. Examples include gender, culture, age, ethnicity, personality traits, and the presence of specific pathologies, among others (cf. Figure \ref{fig:personal-factors}). 
By incorporating individual differences into computational models of affect, researchers can create more personalized and accurate predictions of cognitive and emotional  responses and behaviors. In a human-machine interaction context, the personal factors may pertain to the human or to the machine (system). For instance, human personal factors like age and ethnicity were  proven to influence the individual's affective states \cite{rudovic2018personalized, shaqra2019recognizing}. Similarly, 
gender and personality of a machine (i.e a robot ) were observed to have a direct impact on the affective states of a user \cite{shaqra2019recognizing,salam2016fully}.

\textbf{Gender}. Several studies have addressed the impact of gender on the affective state of users \cite{park2011effects,sidner2004look,pmlr-v173-salam22a,vogt-andre-2006-improving,kim2009towards,kasparova2020inferring}. As an example, a study conducted in the field of HRI \cite{park2011effects} found that the majority of participants favored interacting with robots of the opposite gender. 
Training gender-specific classifiers have been confirmed by multiple works in emotion recognition tasks that it can improve the inference accuracy compared to non-target models in HCI scenarios \cite{vogt-andre-2006-improving, kim2009towards}.

\textbf{Culture}. As noted in the study of \cite{triandis1994culture}, the literature has extensively highlighted variations in social behavior across various societies and cultures.
For example, researchers on emotion recognition have found that considering cultural factors leads to greater accuracy \cite{elfenbein2002universality}. Likewise, an investigation of engagement detection in HRI has uncovered variations in children's behaviors of engagement across different cultural backgrounds \cite{rudovic2017measuring}. These differences have been considered in the development of computational engagement models \cite{rudovic2018culturenet}.

\textbf{Personality}. The literature also demonstrates the impact of personality on affective states. For example, in a triadic HRI study, researchers found a noteworthy relationship between participants' agreeableness and extroversion traits and their perceived enjoyment when interacting with an extroverted robot \cite{celiktutan2015computational}. Furthermore, investigations have delved into the impact of the users' personality on their level of engagement, disregarding the influence of the robot's personality. Results suggest that individuals with higher extroversion scores tend to engage in longer interactions with robots \cite{Ivaldi2015}. %

\textbf{Pathology}.
Pathologies such as Autism Spectrum Disorder (ASD), Bipolar Disorder (BP), and Attention Deficit Hyperactivity Disorder (ADHD) can have an impact on social behavior \cite{rudovic2018personalized, rudovic2018culturenet, rudovic2017measuring, sonuga2013nonpharmacological}. For instance, individuals with ASD may have difficulties with conveying social cues in social interactions \cite{rudovic2018personalized}, while individuals with ADHD may display high-frequent motions and face challenges in maintaining focus \cite{sonuga2013nonpharmacological}. 

Incorporating details about a certain condition or clinical evaluations into models for affective inference of those pathologies offers an improved understanding regarding the reasoning behind these models. However, such approaches are rare in the literature such as proposals implemented to determine a child's affective state within the context of robot-assisted autism therapy \cite{rudovic2018personalized}.

\begin{figure*}
    \centering
    \includegraphics[width=1.1\linewidth]{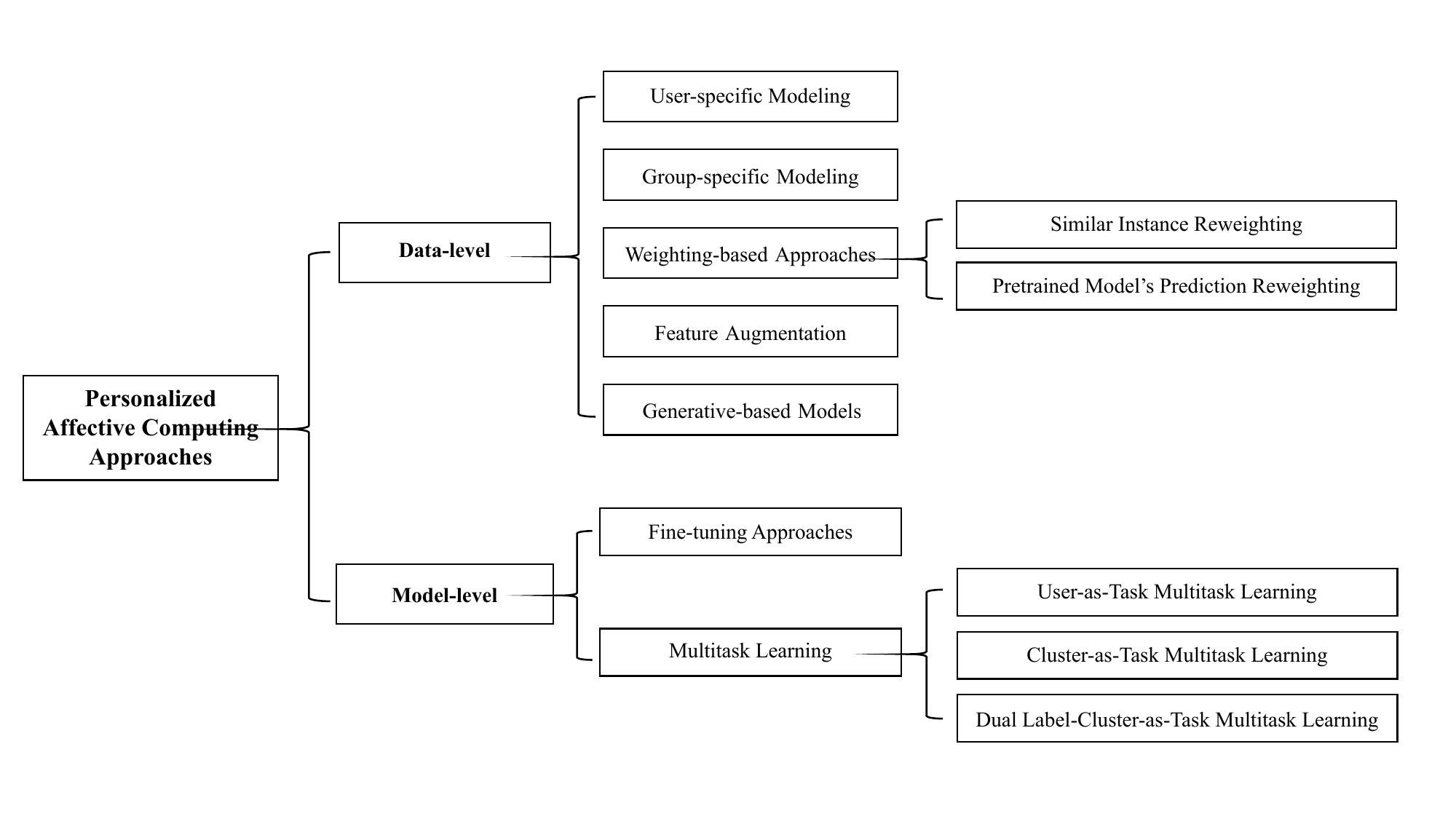}
    \caption{Framework of personalized affective computing techniques reviewed in this paper.}
    \label{fig:Framework}
\end{figure*}
\section{Personalized Affective Computing}
\label{sec:problem-formulation}
\textbf{Problem Formulation \& Definition.} We introduce a formal mathematical model for personalized affective computing, which offers a structured framework to categorize existing personalization techniques and guides the development of future methods.

Let $S$ be a set of individuals $I$. 
Let $D_I$ be the dataset corresponding to each individual $I \in S$.
Let $D_S:=\left\{D_I \mid I \in S\right\}$ be the set of all datasets. The size of $D_S$ is $|S|$, which corresponds to the number of individual datasets. 

A dataset $D_I$ of an individual $I$ consists of a matrix $X$ of input features and a vector or a matrix of outputs $y$, i.e. $D_I:=\{(X, y)\}$. The outputs $y$ can be either real numbers or categorical values (regression vs. classification). 
We define two types of input features: (1) Person-specific ($X_P$) and (2) Non-Person-specific ($X_{NP}$), i.e.  $X:=\{(X_{P}, X_{NP})\}$.  Person-specific features constitute features that describe specific characteristics of the individual such as their gender, age, culture, personality traits, or even typical behavioral clusters. Non-Person-specific features constitute general features relevant to the affective or personality computing task. These may include behavioral features extracted from multimodal data or contextual features extracted from the interaction scenario.

Let $D_G$ be a general dataset that contains the data corresponding to the union of multiple individual datasets. $D_G$ is assumed to be significantly larger than any $D_I$. $D_G$ may or may not be equal to $D_S$, depending on the availability and adequacy of all the individual datasets.

Personalization aims to acquire a model that caters to the specific needs of an individual or a group of individuals with similar characteristics. Consequently, we distinguish two personalization strategies in the literature: (1) user-level, and (2) group-level. 

Methods for personalizing affective computing models differ in whether they make use of the general dataset $D_G$, the individuals datasets $D_I$, or the person-specific features.  The following section details such strategies and approaches. 

\begin{table*}[ht!]
    \scriptsize
    \caption{A summary of the surveyed literature in personalized affective computing. We include target of personalization (User, Group), the context of the study, and the interaction mode (HCI, HHI, HRI).}
    \label{tab:Large Table_data}
    \begin{tabular}{@{}p{0.03\textwidth}p{0.14\textwidth}p{0.04\textwidth}p{0.05\textwidth}p{0.24\textwidth}p{0.24\textwidth}p{0.05\textwidth}@{}}

    \toprule
\multicolumn{1}{l}{\textbf{Level}} & \textbf{Technique} & \textbf{Paper} &  \textbf{Target} & \textbf{Task} & \textbf{Context} & \textbf{Mode} \\ \midrule
\multirow{70}{*}{\rotatebox{90}{\textbf{DATA}}} & \textbf{User-specific} & \cite{daly2018personalised} &  User & Adaptive Music Recommendation & Entertainment & HCI \\
 &  & \cite{cao2015speaker} & User & Emotion Recognition & \begin{tabular}[l]{@{}l@{}} Emotion Understanding\\ \end{tabular} & HCI \\
 &  & \cite{abbasi2009towards} & User & Mental State Recognition & \begin{tabular}[l]{@{}l@{}} Educative\\ \end{tabular} & HHI \\
 &  & \cite{leyzberg2014personalizing} & User & Learner Skills Assesement & \begin{tabular}[l]{@{}l@{}} Educative\end{tabular} & HCI \\
 &  & \cite{shao2021personality} & User & Personality Recognition & \begin{tabular}[l]{@{}l@{}} Social Interaction\end{tabular} & HHI \\
 &  & \cite{shah2021personalized} & User & Depression Recognition & Health & HCI \\
 &  & \cite{ren2022predicting} & User & Emotion Recognition & Emotion Understanding & HCI \\
 &  & \cite{jiang2024personalized} & User & Pain Recognition & Health & HCI \\
 &  & \cite{bangamuarachchi2023inferring} & User & Mood Detection & Health & HCI \\
 &  & \cite{xu2018active} & User & Pain Detection & Health & HCI \\
 &  & \cite{chang2010personalized} & User & Facial Expression Recognition & Facial Expression Understanding & HCI \\
 &  & \cite{chang2009personalized} & User & Facial Expression Recognitionn & Facial Expression Understanding & HCI \\
 &  & \cite{guo2021personalized} & User & Pain detection & Health & HCI \\
 &  & \cite{valenza2014revealing} & User & Emotion Recognition & Emotion Understanding & HCI \\
 &  & \cite{gordon2016affective}& User & Emotion Recognition & Educative & HRI \\
 &  & \cite{janssen2012tune} & User & Emotion Recognition & Entertainment & HCI \\
 &  & \cite{govindarajan2018affective} & User & Emotion Recognition & Health & HCI \\
 &  & \cite{islam2023personalized} & User & Stress Prediction & Health & HCI \\
 &  & \cite{wampfler2022affective} & User & Emotion Recognition & Emotion Understanding & HCI \\
 &  & \cite{li2023temporal} & User & Emotion Recognition & Emotion Understanding & HCI \\
 &  & \cite{sumer2021multimodal} & User & Engagement Detection & Educative & HHI \\
 &  & \cite{john2022personalized} & User & Engagement Detection & Robot Service  & HRI \\
 &  & \cite{apicella2022eeg} & User & Engagement Detection & Educative & HCI \\
 &  & \cite{gupta2020affectivelyvr} & User & Emotion Recognition & Emotion Understanding & HCI \\
 &  & \cite{pradhapan2020role} & User & Emotion Recognition & Emotion Understanding & HCI \\
 &  & \cite{kulic2007affective}& User & Emotion Recognition & Emotion Understanding & HRI \\
 &  & \cite{huang2015identifying}& User & Emotion Recognition & Emotion Understanding & HCI \\
 &  & \cite{yusuf2017individuality} & User & Emotion Recognition & Social Interaction & HCI \\ \cmidrule(l){2-7} 

 & \textbf{Group-specific} & \cite{pmlr-v173-salam22a} &  Group & Personality Recognition & \begin{tabular}[l]{@{}l@{}} Spontaneous Social Interaction\end{tabular} & HHI \\
 &  & \cite{kim2009towards} & Group & Emotion Recognition & Social Interactionn & HCI \\
  &  & \cite{bangamuarachchi2023inferring} & Group & Mood Detection & Health & HCI \\
 &  & \cite{shaqra2019recognizing} & Group & Emotion Recognition & Social Interaction  & HHI \\
 &  & \cite{meegahapola2023generalization} & Group & Mood Detection & Emotion Understanding & HCI \\
 &  & \cite{team2iccv} & Group & Personality Recognition & Educative  & HHI \\
 &  & \cite{kathan2022personalised} & Group & Depression Recognition & Health  & HCI \\
 &  & \cite{tian2021personality} & Group & Emotion Recognition & Emotion Understanding  & HCI \\
 &  & \cite{rukavina2016affective} & Group & Emotion Recognition & Emotion Understanding  & HCI \\
 &  & \cite{kasparova2020inferring} & Group & Engagement Detection & \begin{tabular}[l]{@{}l@{}} Educative \end{tabular} & HRI \\ \cmidrule(l){2-7}

 & \textbf{Weighting-based}  &  \cite{chattopadhyay2012multisource} & User & Fatigue Detection & \begin{tabular}[l]{@{}l@{}} Health\end{tabular} & HCI \\
 &  & \cite{chu2016selective} & User & Facial Expression Detection & Social Interaction & HHI \\
&  & \cite{bangamuarachchi2023inferring} & Group & Mood Detection & Health & HCI \\
 &  & \cite{feffer2018mixture} & User & Affective State Estimation & Educative & HHI \\
 &  & \cite{chu2013selective} & User & Facial Action Unit Detection & No Context & HCI \\ 
 
 &  & \cite{xu2021personalized} & User & Pain Detection & No Context & HCI \\
 &  & \cite{li2019sample} & User & Facial Expression Recognition & Facial Express Understanding& HCI \\
 &  & \cite{fujita2018one} & User & Anomalous Facial Expression Detection & Health & HCI \\
 &  & \cite{wu2011inductive} & User & Emotion Recognition & Emotion Understanding & HCI \\
 &  & \cite{shin2017development} & User & Emotion Recognition & Educative & HCI \\
 &  & \cite{arapakis2010comparison} & User & Emotion Recognition & Emotion Understanding & HCI \\
 &  & \cite{sridhar2022unsupervised} & User & Emotion Recognition & Emotion Understanding & HHI \\
 &  & \cite{bang2018adaptive} & User & Emotion Recognition & Emotion Understanding & HCI \\
&  & \cite{luo2021progressive}& User & Emotion Recognition & Emotion Understanding & HCI \\
 &  & \cite{tran2023personalized} & User & Emotion Recognition & Emotion Understanding & HCI \\ \cmidrule(l){2-7}

 &  \textbf{Generative-based} & \cite{niinuma2022facial} & User & Facial Expression Detection & Social Interaction & HCI \\
 &  & \cite{wang2018personalized} & User & Facial Action Units Recognition & Emotion Understanding & HCI \\
 &  & \cite{yang2018identity} & User & Facial Expression Detection & Social Interaction & HCI \\
 &  & \cite{alyuz2016semi} & User & Engagement Detection & Educative  & HCI \\
 &  & \cite{barros2019personalized} & User & Emotion Recognition & Social Interaction & HCI \\
 &  & \cite{liang2020pose} & User & Facial Expression Detection &Social Interaction & HCI \\
 &  & \cite{salman2022privacy} & User & Facial Expression Recognition & Facial Express Understanding & HHI \\
 &  & \cite{barros2019personalized} & User & Emotion Recognition & Emotion Understanding& HCI \\
 &  & \cite{sun2023exclusive} & User & Emotion Recognition & Emotion Understanding & HCI \\
&  & \cite{garcia2020user} & User & Emotion Recognition & Emotion Understanding & HCI \\
&  & \cite{bang2018adaptive} & User & Emotion Recognition & Emotion Understanding & HCI \\  \cmidrule(l){2-7}

 & \textbf{Feature Augmentation} & \cite{salam2016fully} &  User & Engagement Detection & \begin{tabular}[l]{@{}l@{}} Education \end{tabular} & HRI \\
 &  & \cite{yang2014personalized} & User & Facial Action Units Intensity Estimation & No Context & HCI \\
 &  & \cite{oh2021inductive} & User & Stress Prediction & Health & HCI \\
 &  & \cite{lopez2017personalized} & User & Pain detection & Health & HCI \\
 &  & \cite{vogt-andre-2006-improving} & Group & Emotion Recognition & Social Interaction & HCI \\
 &  & \cite{zhao2018transferring} & Group & Emotion Recognition & Emotion Understanding & HHI \\  \cmidrule(l){2-7} 

  & \textbf{Other} & \cite{triantafyllopoulos2022exploring} & User & Depression Recognition & Emotion Understanding & HCI \\
  \midrule
\end{tabular}
\end{table*}

\begin{table*}[ht!]
    \scriptsize
    \label{tab:Large Table_model}
    \begin{tabular}{@{}p{0.03\textwidth}p{0.14\textwidth}p{0.04\textwidth}p{0.11\textwidth}p{0.21\textwidth}p{0.21\textwidth}p{0.05\textwidth}@{}}
    \toprule
\multicolumn{1}{l}{\textbf{Level}} & \textbf{Technique} & \textbf{Paper} &  \textbf{Target} & \textbf{Task} & \textbf{Context} & \textbf{Mode} \\ \midrule
\multirow{38}{*}{\rotatebox{90}{\textbf{MODEL}}} & \textbf{Fine-tuning} & \cite{kathan2022personalised} &  User & Humor Recognition & \begin{tabular}[l]{@{}l@{}} Social Interaction\\ \end{tabular}& HCI \\

 &  & \cite{Rescigno202035811} & User & Facial Expression Detection & Social Interaction & HCI \\
 &  & \cite{woodward2021towards} & User & Mental Wellbeing Recognition & Emotion Understanding & HCI \\
 &  & \cite{barros2022ciao} & User & Facial Expression Detection & Social Interaction & HCI \\
 &  & \cite{li2020personality} & User & Image Asethetic Detection & Entertainment & HCI \\
 &  & \cite{rudovic2018personalized} & User & Engagement Detection & Educative & HRI \\
 &  & \cite{sridhar2022unsupervised} & User & Emotion Recognition & Emotion Understanding & HHI \\
 &  & \cite{luo2021progressive} & User & Emotion Recognition & Emotion Understanding & HCI \\
 &  & \cite{tran2023personalized} & User & Emotion Recognition & Emotion Understanding & HCI \\
 &  & \cite{zen2016learning} & User & Facial Action Unit Detection & No Context & HCI \\
 &  & \cite{sangineto2014we} & User & Facial Expression Recognition & No Context & HCI \\
 &  & \cite{zen2014unsupervised} & User & Emotion Recognition & Health & HCI \\
 &  & \cite{liu2017deepfacelift} & User & Pain Detection & Health & HCI \\
 &  & \cite{han2020personalized} & User & Pain detection & Health & HRI \\
 &  & \cite{savchenko2022personalized}& User & Facial Expression Recognition & Facial Express Understanding & HCI \\
 &  & \cite{savchenko2022audio} & User & Facial Expression Recognition & Facial Express Understanding & HCI \\
 &  & \cite{zheng2016personalizing} & User & Emotion Recognition & Emotion Understanding & HCI \\
 &  & \cite{chen2017component} & User & Emotion Recognition & Entertainment & HCI \\
 &  & \cite{chen2014linear} & User & Emotion Recognition & Entertainment & HCI \\
 &  & \cite{park2023muse} & User & Emotion Recognition & Emotion Understanding & HHI \\
 &  & \cite{kathan2022personalised} & User \& Group & Depression Recognition & Health& HCI \\
 &  & \cite{gerczuk2022personalised} & User & Depression Recognition & Health & HCI \\
 &  & \cite{gonzalez2021personalizing} & User & Emotion Recognition & Emotion Understanding & HCI \\
 &  & \cite{perz2022personalization} & User & Emotion Recognition & Emotion Understanding & HCI \\
 &  & \cite{sun2023exclusive} & User & Emotion Recognition & Emotion Understanding & HCI \\
 &  & \cite{libman2023ecg} & User & Stress Prediction & Emotion Understanding & HCI \\
 &  & \cite{rudovic2019personalized} & User & Engagement Detection & Educative & HRI \\
 &  & \cite{churaev2022multi} & User & Emotion Recognition & Emotion Understanding & HCI \\
 &  & \cite{garcia2020user}& User & Emotion Recognition & Emotion Understanding& HCI \\\cmidrule(l){2-7} 

 & \textbf{Multitask Learning} & \cite{lopez2018multi} &  Group & Pain Recognition & Health & HCI \\
 &  & \cite{lopez2017multi} & User & Pain Recognition & Health & HCI \\
  &  & \cite{zhao2019personalized} & Group & Emotion Recognition & Emotion Understanding & HCI \\
 &  & \cite{jaques2016multi} & User \& Group & Mood Detection & Health & HCI \\
 
 &  & \cite{taylor2017personalized} & User & Mood Detection & Health & HCI \\
 &  & \cite{saeed2017personalized} & User & Stress Prediction & Health  & HCI \\
 &  & \cite{saeed2018model} & User & Depression Recognition & Health  & HCI \\
 &  & \cite{chithrra2022personalized} & Group & Engagement Detection & Educative   & HRI \\
 &  & \cite{lopez2017physiological} & User & Pain Detection & Health & HCI \\
 &  & \cite{zhao2016predicting} & User & Emotion Recognition & Emotion Understanding & HCI \\
&  & \cite{suman2022investigations}& User & Emotion Recognition & Entertainment & HHI \\
  \cmidrule(l){2-7} 
 & \textbf{Other} & \cite{cai2018inferring} &  Group & Emotion Recognition & Emotion Understanding & HCI \\
\midrule
\end{tabular}
\end{table*}

\textbf{Approaches Taxonomy.} In order to answer our first research question, in the following subsections, we present a comprehensive taxonomy of the existing literature on personalized affective computing methods. Our analysis of the literature has led to a taxonomy which is based on either (1) data-level methods or (2) model-level techniques. We also consider how the different affective contexts have been leveraged with personalization frameworks. 
Table \ref{tab:Large Table_data} presents a summary of the surveyed papers. The surveyed works in personalized affective computing are organized into the  structure  shown in Figure \ref{fig:Framework}.

\subsection{Data-Level Personalization}
Data-level frameworks perform personalization on data before feeding it to the classifier. Techniques that do not expand the target-specific dataset fall under user-specific or group-specific modeling, both of which are types of data-grouping methods. If a technique enlarges the dataset, it's crucial to identify whether this occurs at the instance or feature level. Expansion at the feature level indicates feature augmentation, while at the instance level, it suggests either a generative-based model or a weighting-based technique. Finally, we distinguish generative-based models from weighting-based models based on whether the increased instance comes from a non-target dataset.

\subsubsection{User-specific Modeling}
User-specific modeling is a personalization technique where individual models are uniquely developed and trained for each target user. This approach aims to achieve precise predictions or decisions by tailoring the model to the unique characteristics and behaviors of the user.

\begin{figure}[!ht]
    \center
    \includegraphics[width=\linewidth]{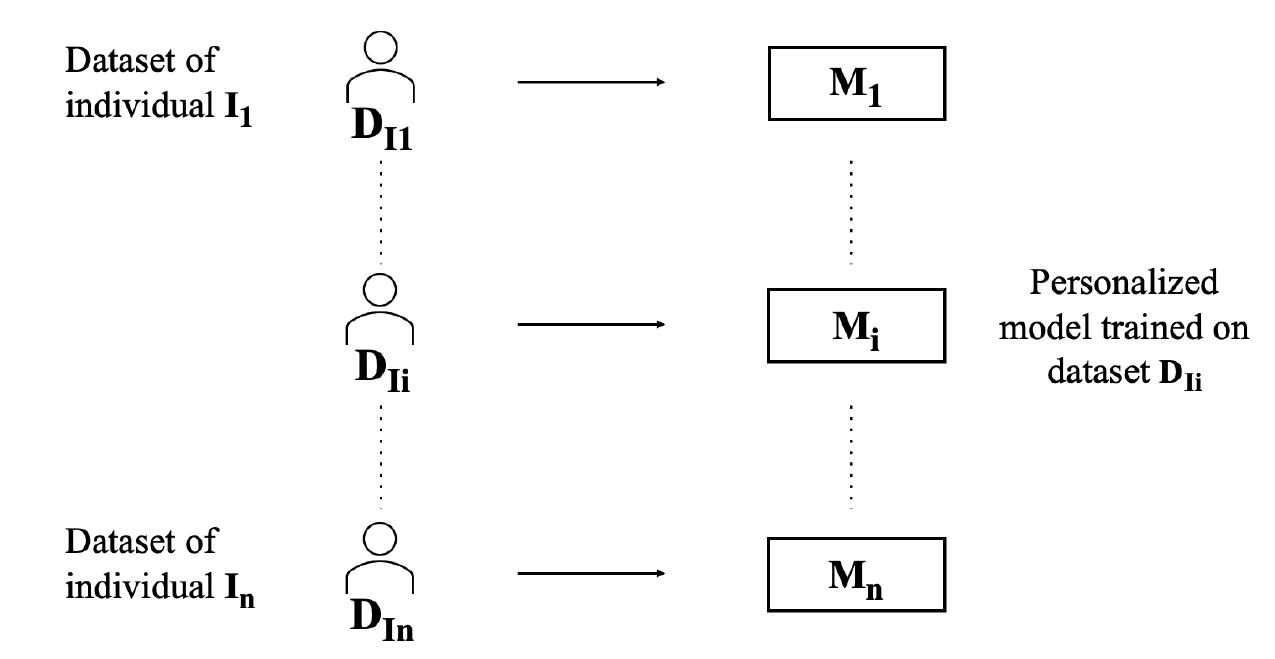}
    \caption{A demonstration of the user-specific modeling method. Each unique model $M_i$ is trained on a dataset $D_{I_{i}}$ corresponding to a unique individual $I_i$.}
    \label{fig:user-specific}
\end{figure}

\textbf{Problem Formulation.} Formally, given $n$ individual datasets $D_{I_{i}}$, each corresponding to an individual $I_i$, where $i \in [1,n]$, a unique model $M_i$ is trained on each $D_{I_{i}}$.  Figure \ref{fig:user-specific} provides an illustration of this approach.

This method does not have a global training stage, i.e. training a generic model on the major dataset (dataset including all the users). This allows the model to excel in improving its predictions and decision-making capabilities based on the target's specific behavior and characteristics. The end result is a model that is highly optimized to the specific needs and requirements of the target user, providing more accurate and relevant results. This approach is particularly advantageous when the user base is limited and the model is intended for extended utilization.

\textbf{Example Approaches.} Examples of automatic user-specific personalized affective models in the literature are as follows: in \cite{daly2018personalised}  a neural network was embedded in the Brain-Computer Music Interface (BCMI) to detect users' affective state from multiple modalities (EEG, peripheral physiological signals, and facial expressions) with the aim to play adaptive music that better matches the user's emotional state. Similarly, a two-layered dynamic Bayesian network was proposed to build an emotionally-personalized computer in \cite{abbasi2009towards}. It uses the images and body gestures of every single student to infer students' mental state (thinking, satisfied, recalling) from their observed behaviors. In a Human-Robot Interaction (HRI) setting \cite{leyzberg2014personalizing}, the robot was endowed with the capacity to assess each student's problem-solving skills as they solved puzzles and customize the order of the lessons accordingly. For depression recognition, a model was proposed to predict individual depression by feeding target data only \cite{shah2021personalized}. In \cite{shao2021personality}, a person-specific personality detection model was formulated. First, they perform a Neural Architecture Search to target a multi-modal CNN structure that best represents the person's facial reactions during a dyadic interaction, using the audio-visual non-verbal input from the partner engaged in the conversation. Then, automatic person-specific personality recognition is achieved by feeding the parameterized CNN model to a graph neural network. A user-level personalized emotion recognition model was also applied through the self-evolving genetic programming (GP) in \cite{yusuf2017individuality} where a user-specific classifier evolves to recognize the person's emotions from their facial action units. In \cite{ren2022predicting}, a personalized ensemble model was developed for each individual separately to predict states of elevated negative affect in adolescents. The privilege in such approach is the grouping of several individualized models at optimal weights to enhance the overall predictive performance for everyone.

\textbf{Discussion.}  User-specific models work well when user interactions are unique and do not generalize, like atypical emotional responses or specialized professional settings \cite{leyzberg2014personalizing}. They also help address privacy concerns with sensitive data that cannot be shared broadly \cite{daly2018personalised}. Additionally, user-specific models suit long-term deployments. In collaborative settings where a user's affect correlates with group dynamics, personalized modeling effectively captures affect \cite{abbasi2009towards}. 

\subsubsection{Group-specific Modeling}

Group-specific modeling is a personalization strategy where models are trained for distinct user groups, categorized by shared profile characteristics such as demographics, preferences, and behavior patterns. This categorization is often achieved through methods like exploratory data analysis or unsupervised clustering, which identify patterns and relationships in the data. Each group receives a uniquely architected model, optimized for its specific traits and behaviors.

\begin{figure}[!ht]
    \center
    \includegraphics[width=\linewidth]{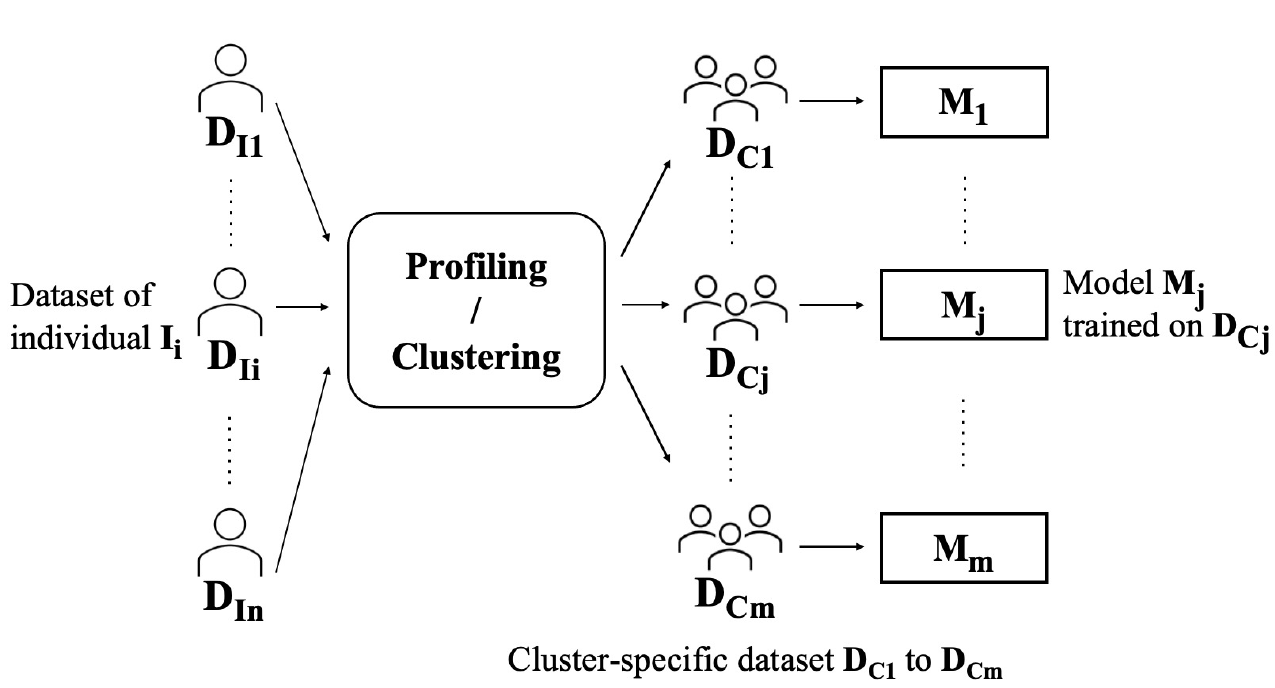}
    \caption{A demonstration of the group-specific modeling method. Each unique model $M_j$ is trained on a cluster dataset $D_{C_{j}}$. Each dataset is formed of a group of datasets $D_{I_{i}}$ corresponding to individuals $I_{i}$ clustered together.}
    \label{fig:Group-Specific-Modeling}
\end{figure}

\textbf{Problem Formulation.} 
Let $S$ be a set of individuals $I$. Let $D_I$ be the dataset corresponding to each individual $I \in S$ and $D_{I_{i}}$ be the dataset of the $i^{th}$ individual included in $S$, where $i \in [1,n]$. Given that $D_S:=\left\{D_I \mid I \in S\right\}$, $D_S$ is further subdivided into $m$ clusters resulting in the datasets $D_{C_{j}}$, where $j \in [1,m]$, by grouping individual datasets $D_{I_{i}}$ together. 
A unique model $M_j$ is then trained on each cluster dataset $D_{C_{j}}$. 
If person-specific features $X_P$ are available, they can be used to obtain the different clusters. Otherwise, $X_{NP}$ are used to cluster the users based on behavioural patterns or other characteristics. A graph representation is provided in Figure \ref{fig:Group-Specific-Modeling}.

When making predictions, the appropriate model is selected based on the user's group membership. This not only allows the system to provide highly personalized results that are relevant but also ensures the model's fairness, i.e., treating similar individuals equivalently, as it allows advanced modeling and fine-tuning that is catered to the specific needs of each target group. 

\textbf{Example Approaches.} A foundation example of group-specific personalization is presented in \cite{pmlr-v173-salam22a}, where personalized personality recognition models were trained separately for different genders. While operating on a shared feature space of visual, textual, and audio traits, optimized neural architectures were derived per group via Neural Architecture Search to accentuate intrinsic differences. Similarly, \cite{kim2009towards} developed vocal emotion classifiers tailored to age and gender groups. Extending this approach, \cite{meegahapola2023generalization} created nation-specific mood inference models, indicating the effectiveness of cultural customization. Complementing these parallel models, \cite{shaqra2019recognizing} pioneered a sequential, two-stage architecture by first predicting subject age and gender to route inputs to the appropriately specialized emotion classifier. These studies affirm the use of broader demographic groupings to enable more personalized systems in the near term, before sizable enough datasets exist to customize models for every individual user. In \cite{kasparova2020inferring}, a Team-LSTM model was proposed for each student group besides its student-specific individual LSTM model, both of which are later combined to process visual cues and output engagement prediction for every single student. 

\textbf{Discussion.} In a collaborative setting where the user's affective states are highly correlated with the group work condition \cite{kasparova2020inferring}, the group-specific modeling technique is an ideal choice for developing computational models of affect. 

\subsubsection{Weighting-based Approaches}

The weighting-based personalization approach involves re-weighting training samples or person-specific model predictions from similar users' datasets based on their similarity with the target user's samples. This approach tries to boost the performance of the target user's prediction task, especially when there is a lack of target-specific data available for training. By adjusting the weights of the samples, the model can better capture the unique characteristics of the target user, resulting in more accurate predictions.

The weighting framework typically comprises two components. Firstly, a similarity measurement or distance metric is employed to estimate the relevance between the target-specific dataset and the general users' dataset. Secondly, the calculated relevance is assigned as weights to each general instance or model prediction. In this manner, instances with higher relevance are more likely to be selected and combined with the target user's data to train the personalized model, while classifiers with higher relevance typically impact greater on predicting the affective states of the target user.

\begin{figure}[!ht]
    \center
    \includegraphics[width=\linewidth]{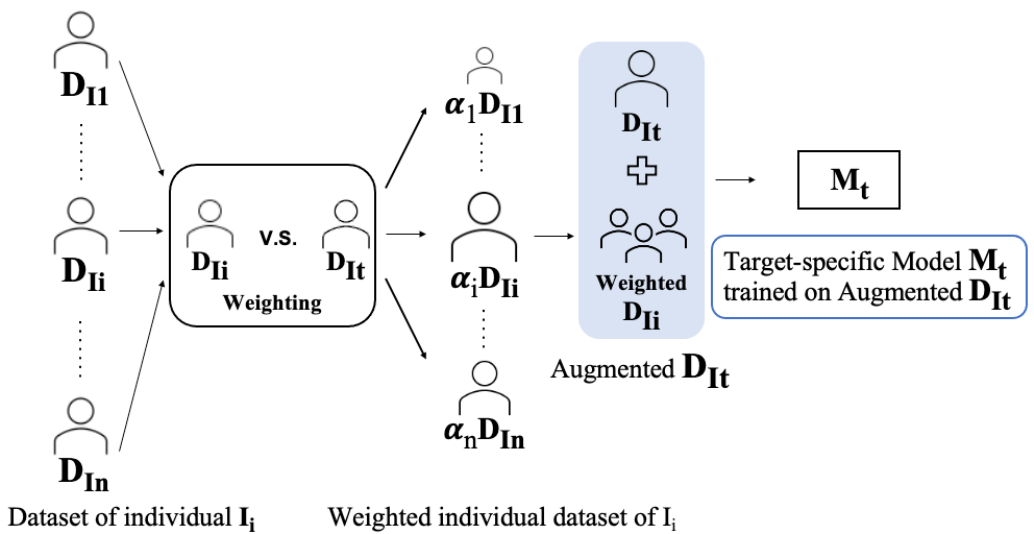}\\
    (a) Similar instances re-weighting.
        \includegraphics[width=\linewidth]{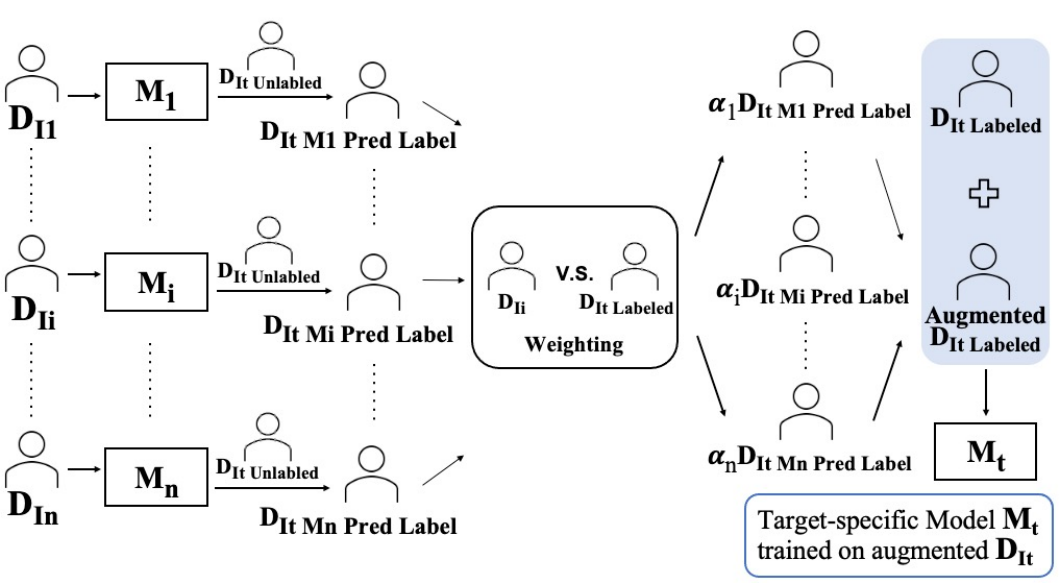}\\
        (b) Person-specific models predictions re-weighting.
    \caption{A demonstration of the weighting-based approach. In (a), a distance metric $\alpha_i$ is calculated between each individual dataset $D_{I_i}$ and the target user dataset $D_{I_t}$. All the resulting weighted datasets $\alpha_i D_{I_i}$ are grouped together to form an augmented dataset used to train the final model $M_t$. In (b), individual models $M_i$ are trained on each $D_{I_{i}}$. Because of the unlabeled data provided, each $M_i$ makes a prediction $D_{I_{t}M_{i}Pred Label}$ that is multiplied by a confidence estimate $\alpha_i$ to obtain the Augmented $D_{I_{t} Labeled}$. Finally, the Augmented $D_{I_{t} Labeled}$ and the originally labeled target data $D_{I_{t} Labeled}$ are combined together to train the individual-specific model $M_t$.}
    \label{fig:Weighting-BasedApproaches}
\end{figure}

\textbf{Problem Formulation.} A weighting-based approach that re-weights similar instances can be formulated as follows. 
Let $S$ be a set of individuals $I$. 
Let $D_I$ be the dataset corresponding to each individual $I \in S$ and $D_{I_{i}}$ be the dataset of the $i^{th}$ individual included in $S$, where $i \in [1,n]$. Meanwhile, let $D_{I_{t}}$ be the dataset of the target user. A similarity estimate $\alpha_i$ between each $D_{I_i}$ and $D_{I_t}$ is calculated according to a selected distance metric. The derived similarity value $\alpha_i$ is then assigned to each corresponding $D_{I_i}$ to get the weighted individual dataset $\alpha_n D_{I_i}$. All the weighted individual datasets $\alpha_i D_{I_i}$ are grouped to train the target-specific model $M_t$ and perform future affective tasks on the target user. Figure \ref{fig:Weighting-BasedApproaches} (a) shows an illustration of this approach.
An example work that uses this data-level re-weighting framework is that of \cite{chu2016selective}, a Selective Transfer Machine that combines the two-step approach was introduced for action unit and expression detection. It simultaneously trains the parameters of the classifier and re-weights the training samples that have the highest relevance to the test user. 

Another weighting-based framework for personalization estimated the weighting of person-specific model predictions rather than data instances. Figure \ref{fig:Weighting-BasedApproaches} (b) provides an illustration of the technique. Let $D_{I}$ be the dataset for an individual $I$ within a population $S$. $D_{I_{i}}$ is the dataset for the $i^{th}$ individual in $S$, where $i \in [1,n]$. $D_{I_{t}}$ is the dataset of a target individual. Within $D_{I_{t}}$, there are labeled data $D_{I_{t} Labeled}$ and unlabeled data $D_{I_{t} Unlabeled}$. The weighting mechanism helps to turn $D_{I_{t} Unlabeled}$ into labeled data to facilitate training the model $M_t$ under scenarios where there is no sufficient $D_{I_{t} Labeled}$. The framework begins by training $n$ individual models $M_i$ on each $D_{I_{i}}$. When an unlabeled target input $D_{I_{t} Unlabeled}$ is provided, each $M_i$ makes a prediction, denoted as $D_{I_{t}M_{i}Pred Label}$. Next, a confidence estimate $\alpha_i$ is calculated for each $M_i$ (e.g. using a gating network). The weights $\alpha_i$ are multiplied by each $D_{I_{t}M_{i}Pred Label}$ to obtain the Augmented $D_{I_{t} Labeled}$. Finally, the Augmented $D_{I_{t} Labeled}$ is combined with the originally labeled target data $D_{I_{t} Labeled}$ to train the individual-specific model $M_t$.

\textbf{Example Approaches.} In \cite{chattopadhyay2012multisource}, the proposed personalized fatigue detection scheme builds on the weighting framework. Firstly, it trains one fatigue classifier for each user in the general dataset. Then, assuming that some users in the general dataset share a similar distribution with the target user, it assigns "pseudo-labels" for the unlabeled target user-specific data using an estimated target classifier, which is derived from a weighted combination of all the pre-trained source domain classifiers, namely, the classifiers for each user in the general dataset. Then, the target-specific model is trained using both the pre-labeled and pseudo-labeled target-specific samples for future predictions.

Similarly, the study by \cite{feffer2018mixture} introduced a unique gating mechanism for training individual expert models, allowing effective adaptation to new subjects with limited personal data. Initially, the framework employed a CNN to pre-train facial feature extractors using a comprehensive facial expression dataset. It then fine-tuned multiple expert sub-networks, each corresponding to a different subject in the training dataset, using subject-specific data. In the testing phase, a gating network was used to determine the optimal contribution of each expert sub-network based on the input data. The final subject-specific models for testing were created by integrating the pre-trained feature extractor with the outputs from various expert sub-networks, blended according to the determined weightings.

\textbf{Discussion -- } Weighting-based approaches can improve target user prediction by leveraging related source domain data when target user data is insufficient. This method works well when: (1) target user data is limited; (2) similar users' datasets exist; or (3) direct data augmentation is impossible. However, properly selecting source domain instances or models is critical for success.

\subsubsection{Generative-based Models} 
Generative-based personalization models follow a two-step approach to perform personalized predictions. The driven idea behind this approach is to overcome the limitations of having insufficient target data in a given dataset with artificially generated samples.

In the first stage, data of the target user is generated using a data generation approach (e.g., using an adversarial model). In the second stage, the classifier, which is tasked with making the affective state inference, is trained based on the combination of the newly generated target-specific data and the original target-specific data. 

\begin{figure}[!ht]
    \center
    \includegraphics[width=\linewidth]{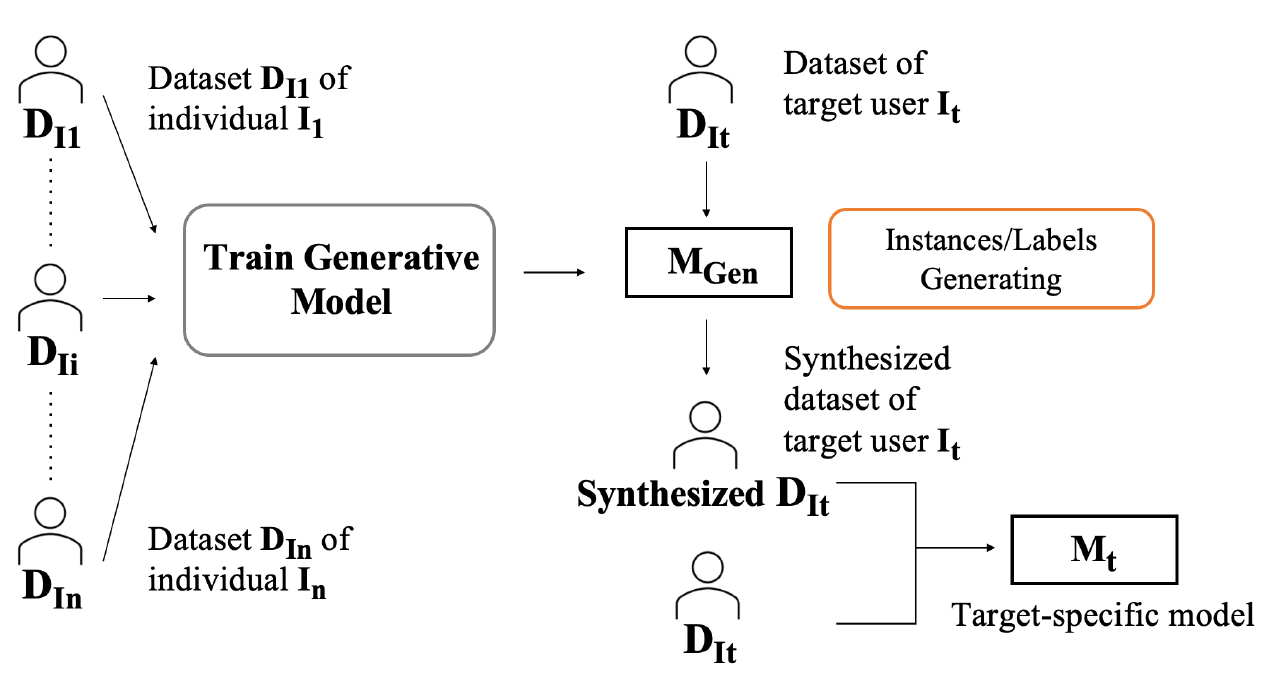}
    \caption{A demonstration of the generative-based model. All individual datasets $D_{I_i}$ are used to train the generative model $M_{Gen}$. $M_{Gen}$ is then applied to the target dataset $D_{I_t}$ by generating labels and instances to produce the synthesized $D_{I_t}$. Finally, the original target dataset $D_{I_t}$ and the synthesized $D_{I_t}$ are combined together to train the target-specific model $M_t$. }
    \label{fig: GenerativeModels}
\end{figure}

\textbf{Problem Formulation.} As pictured in Figure \ref{fig: GenerativeModels}: let $S$ be a set of individuals $I$. Let $D_I$ be the dataset corresponding to each individual $I \in S$ and $D_{I_i}$ be the dataset of the $i^{th}$ individual included in $S$, where $i \in [1,n]$. $D_S:=\left\{D_I \mid I \in S\right\}$. And let $D_{I_t}$ be the dataset of the target user.

Firstly, all $D_I$ are used to train the generative model $M_{Gen}$, which normally captures the patterns in $D_S$ and learns to synthesize new instances or labels for any individual belonging to $S$. Given some labeled/un-labeled data of the target individual $D_{I_t}$, the generative model $M_{Gen}$ enlarges the target-specific dataset $D_{I_t}$ by either generating new instances for $I$ or labeling the un-labeled data of $I$. Then, the original target dataset $D_{I_t}$ combined with the enlarged $D_{I_t}$ are tasked with training the target-specific model $M_t$.

\textbf{Example approaches.} Previous work \cite{niinuma2022facial} adapted a Generative Adversarial Network (GAN)-based model to generate synthetic images with the facial actions unit of the target user. The novel images not only help produce an AU-balanced dataset but also enable the personalization of pre-trained networks, allowing further fine-tuning of the generic models. In the same manner, \cite{wang2018personalized} facilitates the personalized facial action units recognition by jointly learning a GAN-based network and a subject-independent Action Unit Classifier for each target user. In \cite{yang2018identity}, the proposed identity-adaptive learning algorithm improves the facial expression recognition performance by training multiple models for each facial expression. The GAN network then synthesizes images of the different facial expressions of the target user. Together with the former user-specific facial expression dataset, the generated images will be used to train a person-specific classifier for each target user. 

In \cite{alyuz2016semi}, a semi-supervised hierarchical structure is used to generate labels for user-specific data to facilitate final prediction. It first initialized the generic random forest classifiers for appearance (i.e. pose, facial expression)  and context modalities (i.e. user gender, session information, task performance). Then, the context classifier was calibrated first so that it can be used to automatically label data for the appearance classifier. 

In \cite{barros2019personalized}, a generative-adversarial autoencoder model was tasked with generating target-specific images for personalized online emotion recognition. It contains a Prior-Knowledge adversarial autoencoder (PK) and a set of discriminators, which will learn the prior knowledge of generic emotional representation. The model will synthesize the faces of every single targeted person to form the personalized affective memory, which will be used to train the Growing-When-Required Networks (GWR) to perform individual-level inference. In the same manner, \cite{liang2020pose} suggested an encoder-decoder architecture to learn the pose and expression-related feature representations from the universal dataset, which will collaborate with corresponding discriminators to produce user-specific face images using the encoded features from the target's original images. Then, the personalized facial expression classifier is trained on the combined synthetic images with the original ones.

\textbf{Discussion.} The primary benefit of generative-based models lies in its ability to achieve high-accuracy personalization, even with scarce target data. This is accomplished by generating new data that closely represents the user of interest, which is then used to train a personalized model. However, it's important to note that the generation of unrealistic synthetic images by the adversarial-based model could potentially hinder the effectiveness of classifier training.

\subsubsection{Feature Augmentation}

Feature augmentation focuses on creating new features from an existing affective dataset to enhance the accuracy of personalized predictions. This method utilizes known correlations between different types of affective states, such as personality and emotion expression, to generate extra user-specific features. For instance, a user's engagement behavior might correlate with their personality traits. By leveraging such correlations, it's possible to produce additional features unique to an individual user. After generating these user-specific features, they are integrated with the existing dataset features. The enriched feature set is then employed to train a generic model, which subsequently makes user-specific predictions by incorporating personal characteristics through these features.

\begin{figure}[!ht]
    \center
    \includegraphics[width=\linewidth]{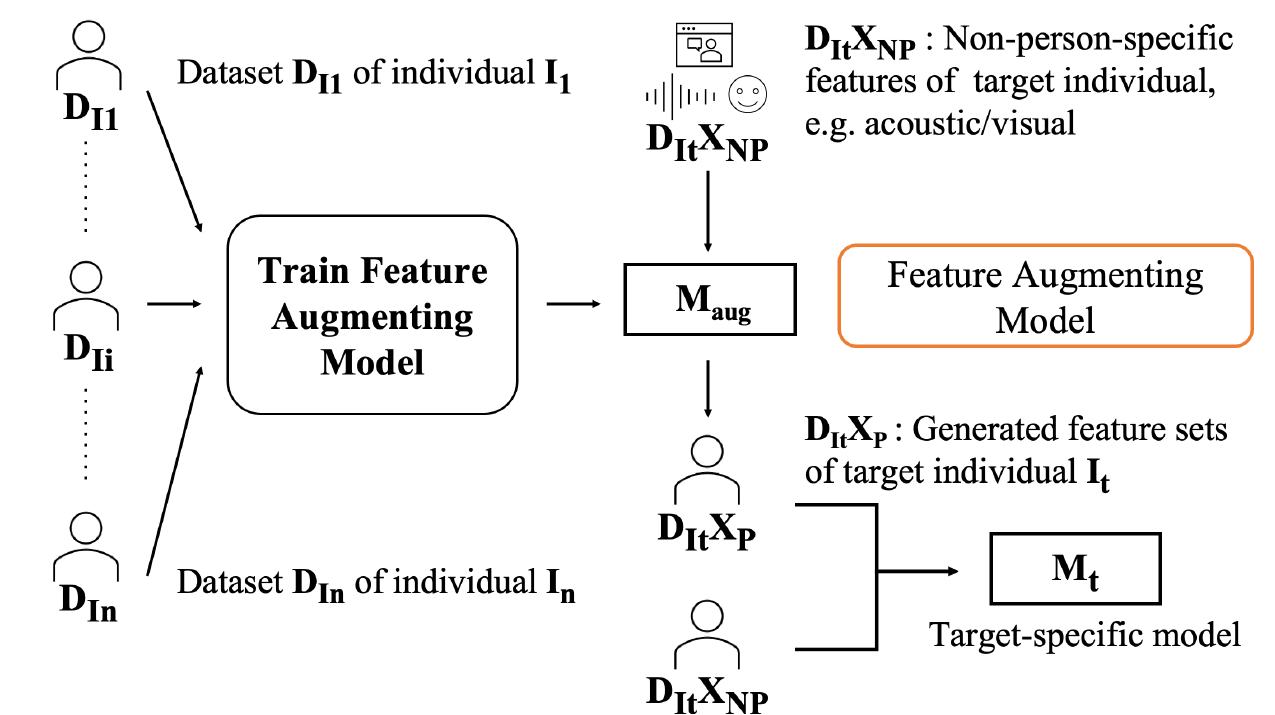}
    \caption{A demonstration of the feature augmentation approach. All individual datasets $D_{I_i}$ are used to train a feature augmenting model $M_{aug}$ that learns to predict person-specific features $X_{P}$ from non-person-specific features $X_{NP}$. Later, $D_{It} X_{P}$ and $D_{It} X_{NP}$ are combined together to train the target-specific affective model $M_t$.}
    \label{fig:FeatureAugmentation}
\end{figure}

\textbf{Problem Formulation.} The feature augmentation mechanism is presented in Figure \ref{fig:FeatureAugmentation}. Let $S$ be a set of individuals $I$. Let $D_I$ be the dataset corresponding to each individual $I \in S$ and $D_{I_i}$ be the dataset of the $i^{th}$ individual included in S. And let $D_{I_n} X_P$ and $D_{I_n} X_{NP}$ represent the person-specific and non-person-specific features of the input for individual $n$, respectively.

Given $D_S:=\left\{D_I \mid I \in S\right\}$ where each $D_I$ consists of two types of features: $X_{NP}$, the non-person specific individual features, like normal acoustic or visual features; and $X_{P}$, the person-specific individual features, like gender, age, personality etc. After training, a feature augmenting model $M_{aug}$ learns to predict $X_{P}$ from $X_{NP}$. The feature augmenting model $M_{aug}$ will be applied on $D_{It} X_{NP} $ to generate predicted $D_{It} X_{P}$, which will later be combined with the original non-person-specific individual features $D_{It} X_{NP}$ to train the target-specific affective model $M_t$.

\textbf{Example Approaches.} As an example, \cite{vogt-andre-2006-improving} proposes a framework to improve emotion recognition from speech by combining automatic gender detection with gender-specific classifiers. The process begins with the automatic identification of gender through acoustic features that extend beyond pitch analysis. Then, the detected gender is used to select between a male or female emotion classifier, which are trained on same-gender utterances only. This combined system outperforms the gender-independent classifier by a 2-4\% accuracy margin across two distinct emotional speech datasets. 

The work of \cite{salam2016fully} uses the generated personality predictions as features for later engagement detection. In \cite{zhao2018transferring}, a hierarchical feature augmentation framework for emotion recognition was proposed. The framework is called FN-MTL and includes two models: the main emotion recognition model and the sub-models for user profile (i.e., age, gender). At the first level of the hierarchical structure, the sub-models learn to generate age and gender attributes from the input features and will be fed straight away to the high-level neural network of the main model. At the second level of the hierarchical structure, the main model takes user profile information as high-level attributes of emotion and performs emotion recognition. The personalization technique helps capture the complex relationship between user profile attributes with emotions. 

\textbf{Discussion.} The benefit of this technique is that it leverages the correlations between different affective state types and personal factors (e.g., age, gender, etc.) to generate additional user-specific features. This, in turn, results in a more nuanced understanding of a user's affective state, which leads to improved personalized predictions.
It's important to note that feature augmentation can be applied in any domain where correlations between different features can be exploited to generate additional information about a user. In addition, feature augmentation can be combined with other personalization techniques, such as adversarial-based personalization, for even more accurate personalized predictions.
 
\subsubsection{Other}
According to our categorization criteria, the personalization approaches proposed by \cite{triantafyllopoulos2022exploring} do not fit into any of the existing categories. Therefore, in this section, we shortly describe the method used in this study. In \cite{triantafyllopoulos2022exploring}, two separate encoder networks were trained separately for two purposes - emotional feature extraction and encoding unlabeled samples from the target speaker respectively. Then the output of the second encoder, which encodes the unlabeled target speaker samples, is used to select relevant features from the first encoder via dot-product attention. This adapts the model to the target speaker. Additionally, a speaker and an emotion classifier are attached to each encoder. In this way, the proposed method avoids the need for clearly labeled target samples and makes personalized predictions for specific speakers.

\subsection{Model-Level personalization}
The model-level frameworks, on the other hand, incorporate personalization in their classifier. They can be categorized into two approaches depending on whether they minimize the combination of multiple loss functions: (1) Fine-tuning and (2) Multitask Learning. Multitask learning involves minimizing the combination of multiple loss functions at a time, while fine-tuning approaches do not.

\subsubsection{Fine-tuning Approaches}

The fine-tuning approach is by far the most widely adapted in personalized affective computing. It assumes that the affective inference model for a specific user and the larger group where it belongs share some parameters or prior distribution of hyperparameters. This approach is particularly effective when the target user and its original group are closely related but with only limited labeled data for the target user.

To train a personalized model for the target user, the pre-trained model on the larger group where the target belongs (source task) is used as the initialization for a personalized model of the target user (target task). The pre-trained model's shared parameters can be fine-tuned to fit the target user, either by freezing some layers in the pre-trained model or by using a smaller learning rate for the pre-trained parameters. The core idea lies in the fact that the pre-trained model on the larger group has learned some useful representations or features that can be transferred to the target user model, and by fine-tuning the pre-trained parameters on the target task (target user state), the model can better capture the unique characteristics of that particular user.

\begin{figure}[!ht]
    \center
    \includegraphics[width=\linewidth]{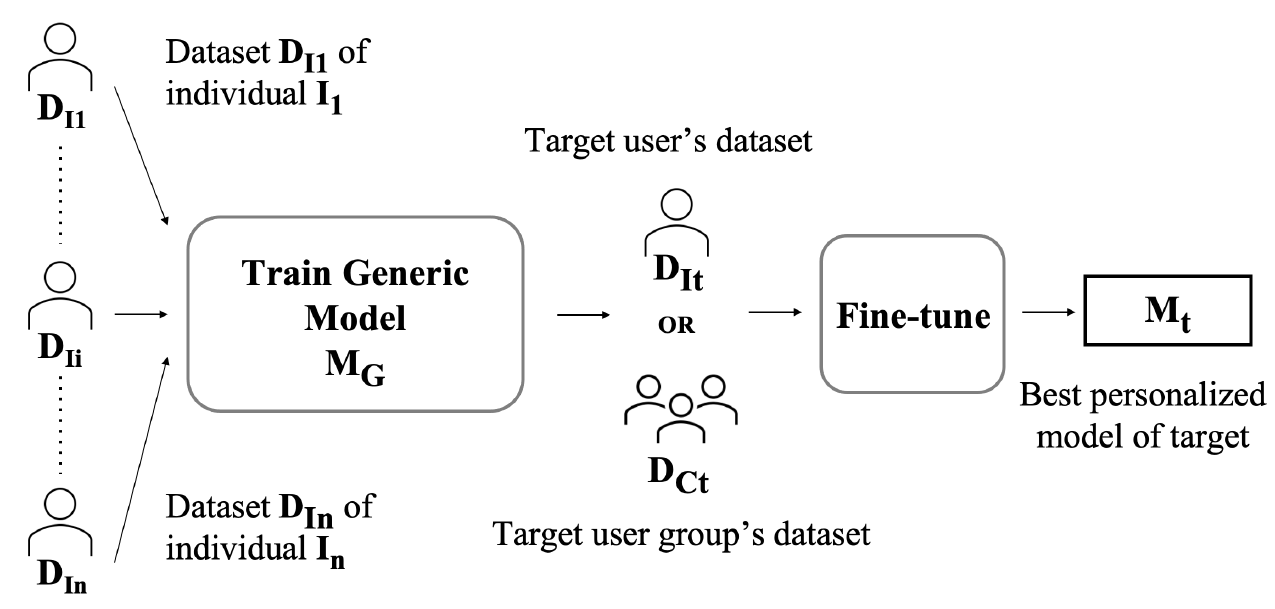}
    \caption{A demonstration of the fine-tuning approach. All individual datasets $D_I$ excluding the target user dataset $D_{I_t}$ and the target user group dataset $D_{C_t}$ are used to train a generic model $M_{G}$. $M_{G}$ is then applied to either $D_{I_t}$ or $D_{C_t}$ while fine-tuning processes are executed to generate the best-performing personalized model $M_t$.}
    \label{fig:Fine-tuningApproaches}
\end{figure}

\textbf{Problem Formulation.} Figure \ref{fig:Fine-tuningApproaches} illustrates the fine-tuning approach. Let $S$ be a set of individuals $I$. 
Let $D_I$ be the dataset corresponding to each individual $I \in S$ and $D_{I_i}$ be the dataset of the $i^{th}$ individual included in $S$, where $i \in [1,n]$. Meanwhile, let $D_{I_t}$ and $D_{C_t}$ be the dataset of the target user and the target user group, respectively. Given $D_S:=\left\{D_I \mid I \in S\right\}$ the set of all individual datasets $D_I$ except for $D_{I_t}$ or $D_{C_t}$, a generic model $M_{G}$ is trained first to capture the ideal starting values of the model's parameters. Next, the dataset from the target user $D_{I_t}$ or the target user group $D_{C_t}$ is fed to the pre-trained models $M_{G}$ while fine-tuning processes are executed to derive the best-performing personalized model $M_t$.

\textbf{Example Approaches.} For example, \cite{kathan2022personalised} proposed fine-tuning the pre-trained general model for individual-level humor recognition by manually fine-tuning for individual humor inference tasks. In another study \cite{Rescigno202035811}, researchers employed a widely-used CNN architecture known as AlexNet. This architecture was initially trained on the AffectNet dataset, serving as a capable feature extractor. The network was then divided into two parts: the convolutional layers and the dense layers. While the convolutional layers were kept unchanged, the dense layers were fine-tuned using a smaller dataset specific to each subject. During this fine-tuning process, the convolutional layer weights remained fixed, ensuring that the previously learned features were retained, while the dense layers were adjusted to the individual subject's data.

A similar approach was taken in \cite{woodward2021towards}, which aimed to create personalized models for detecting users' mental well-being states. They began by training a CNN layer for feature extraction using a general dataset, after which this layer's weights were frozen. The frozen 1D CNN layer was then combined with new fully connected layers, resulting in a trainable model. This combined architecture was fine-tuned on specific users' datasets, allowing the model to adapt to the characteristics of each user's data and enhance its performance in detecting mental well-being states.

The proposed personalization architecture in \cite{barros2022ciao} also leveraged the feature extraction ability of the pre-trained convolutional encoder in facilitating non-universal facial expression recognition. It attached a new convolutional layer to the encoder aiming to learn the target dataset's specific representation. In \cite{li2020personality}, an inter-task fusion was suggested to fine-tune the generic aesthetics assessment model for personalized image aesthetic prediction. Proven that personality traits largely determine people's aesthetic preference for an image, the work proposed to derive the individual aesthetic score of an image as a sum of the image's generic aesthetic score and the linear combination individual's personality traits and that of the image. During the personalization stage, a small group of individual-rated images is used to train the generic aesthetic assessment model. 

Besides the individual-level fine-tuning, \cite{rudovic2018personalized} also proposed a group-level fine-tuning framework in detecting children's affect and engagement level in robot-assisted autism therapy. A deep neural network architecture called PPA-net was designed to incorporate the child's demographic (culture, gender) and contextual information (external experts' autism annotations) to account for individual differences. Specifically, the model architecture has separate branches nested by culture (Asia/Europe) and gender, allowing layers to be personalized based on broad demographics differences.   through the personalized layers and training strategies. This contextual information and model architecture enable more personalized perception. 

In \cite{rudovic2019personalized}, a human-assisted fine-tuning framework was introduced. To accurately perceive user-specific engagement levels, it incorporates human efforts for relabeling to ensure faster convergence of the model. During the personalization, the trained generic engagement classifier will task with providing primary inference on individual user videos. Among all the primary inferences, those with less distinguishability are queried to label by a human expert offline. Since the model re-learns from these human-labeled data, the personalized model is derived in a faster and more robust manner. 

With the feature space sharing method, the model attempts to identify a shared feature space between the source and target domains. This enables the model to map the instances in the source domain to the target user data in a more straightforward manner. This is done by training a shared feature extractor on both domains, which can then be used to extract relevant features from both the source and target data. A sample feature space sharing is \cite{Rescigno202035811}, the pre-trained feature extraction convolutional layer for the generic dataset is frozen and directly re-applied to extract features of the target dataset. 

\textbf{Discussion.} The fine-tuning approach can be an effective technique for boosting the performance of the target task, especially when the source and target tasks share similar characteristics. However, it is important to carefully select the pre-trained model and hyperparameters to ensure that they are appropriate for the target task and to avoid overfitting to the source task.

\subsubsection{Multitask Learning Approaches}
Multitask learning \cite{caruana1993multitask}, is also widely adapted in personalizing affective computing models. In multitask learning, a single model is trained to perform multiple related tasks simultaneously. The model benefits from shared information and parameters across these tasks, enforced by similarity constraints, leading to more efficient and effective learning than training each task in isolation.

Typically, such a model includes multiple output layers, with each layer dedicated to a different task. During the training process, the model's parameters are adjusted to minimize the aggregate loss across all tasks, rather than focusing on the loss for each individual task. This joint optimization enables the model to discern both common patterns and features applicable to all tasks, as well as unique, task-specific characteristics.

Existing works in affective computing using multitask learning are developed in four types of scenarios: 1) \textit{Label-as-Task}: treat each affective computing task (e.g. engagement inference) as one task, while treating another affective computing task (e.g. personality classification) as another \cite{lopez2018multi,jaques2016multi,taylor2017personalized} ;  2) \textit{User-as-Task}: consider predicting the affective state of a single user as a task, and predicting that of another individual user at the same time \cite{jaques2016multi,saeed2017personalized,saeed2018model,lopez2017multi} ; 3) \textit{Cluster-as-Task}: let predicting the affective state of a cluster of users as one task and predict that of another cluster in parallel \cite{jaques2016multi,lopez2018multi} and 4) \textit{Dual Label-Cluster-as-Tasks}: treat predicting the affective state of a cluster as one task and the labeling of the clusters as another \cite{chithrra2022personalized}. The latter three scenarios are widely adapted to personalization. In the literature, various techniques have been developed to personalize affective computing models for the three scenarios above. 

\textbf{Problem Formulation.} In multitask learning, we also present the following formulation: let $S$ be a set of individuals $I$. Let $D_I$ be the dataset corresponding to each individual $I \in S$ and $D_{I_i}$ be the dataset of the $i^{th}$ individual included in $S$ where $i \in [1,n]$. Let $D_{I_t}$ and $D_{C_t}$ be the dataset of the target user or the target user group, respectively. 

\begin{figure}[!ht]
    \center
    \includegraphics[width=\linewidth]{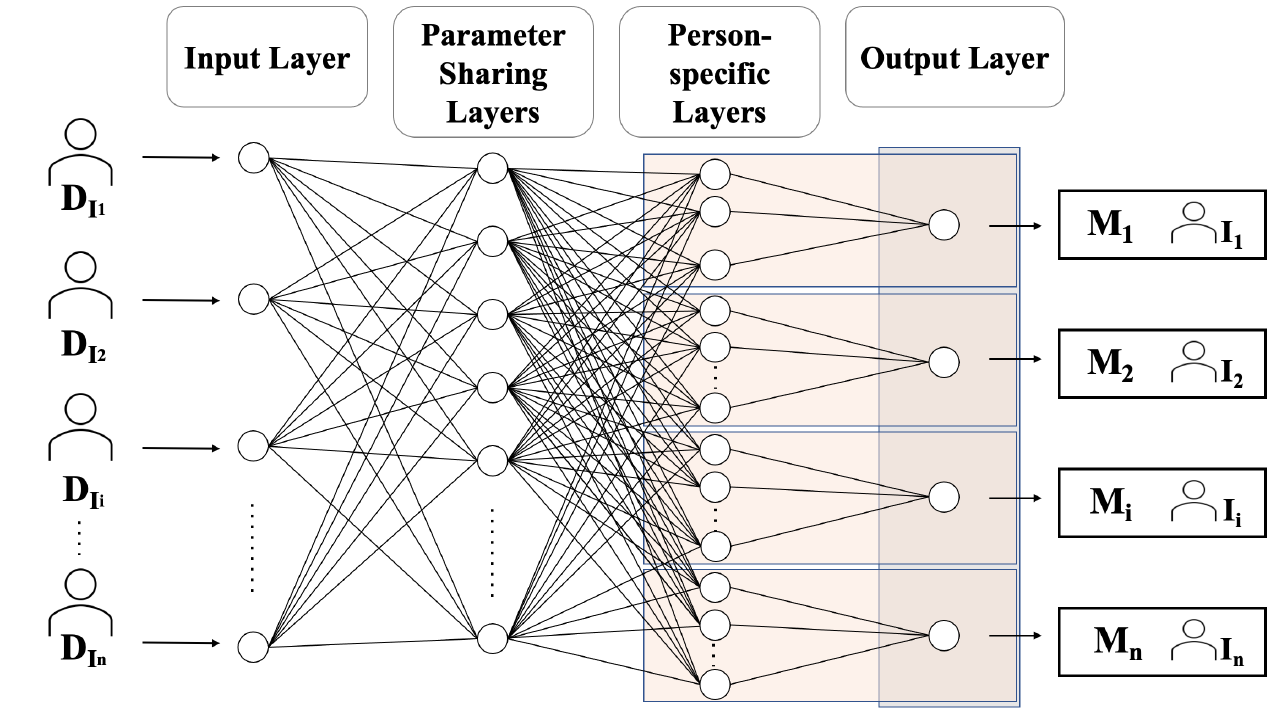}
    \caption{A demonstration of the User-as-Task multitask learning approach. All individual datasets $D_{I_1} \dots D_{I_n}$, along with their corresponding models ${M_1 \dots M_n}$  are trained simultaneously by optimizing a loss function across all individuals $I_1 \dots I_n$.}
    \label{fig: User-as-Task}
\end{figure}

In User-as-Task multitask learning (cf. Figure \ref{fig: User-as-Task}), given $D_S:=\left\{D_I \mid I \in S\right\}$ the set of all individual datasets, corresponding individual-specific models ${M_1, M_2, \dots, M_i, \dots,Mn}$ are trained simultaneously by optimizing an aggregated loss function across all individuals $I_1, I_2, \dots, I_i, \dots, I_n$. 

\begin{figure}[!ht]
    \center
    \includegraphics[width=\linewidth]{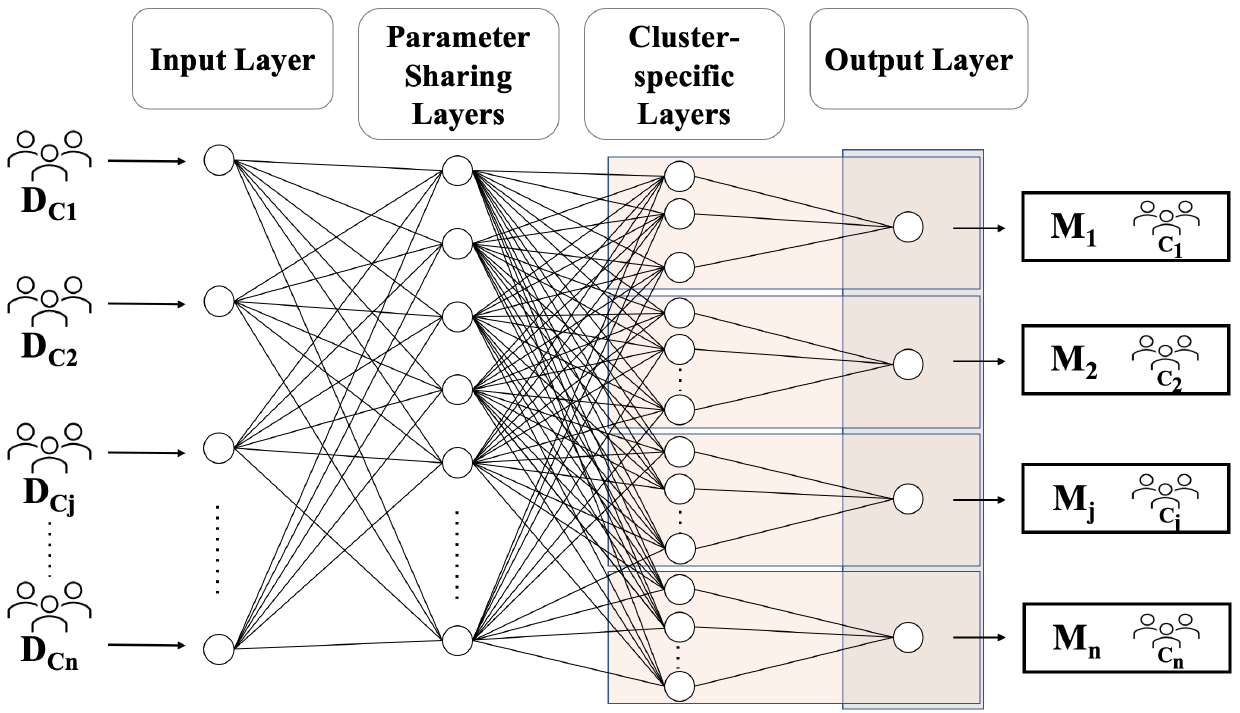}
    \caption{A demonstration of the Cluster-as-Task multitask learning approach. All cluster datasets $D_{C_1} \dots D_{C_n}$, along with their corresponding models ${M_1 \dots M_n}$  are trained simultaneously by optimizing a loss function across all user clusters $C_1 \dots C_n$.}
    \label{fig: Cluster-as-Task}
\end{figure}

In Cluster-as-Task multitask Learning (cf. Figure \ref{fig: Cluster-as-Task}), given $D_S:=\left\{D_C \mid C \in S\right\}$ the set of all cluster datasets, corresponding cluster-specific models ${M_1, M_2, \dots, M_i, \dots ,Mn}$ are trained simultaneously by optimizing an aggregated loss function across all user clusters $C_1, C_2, \dots, C_j, \dots, C_n$. 

\begin{figure}[!ht]
    \center
    \includegraphics[width=\linewidth]{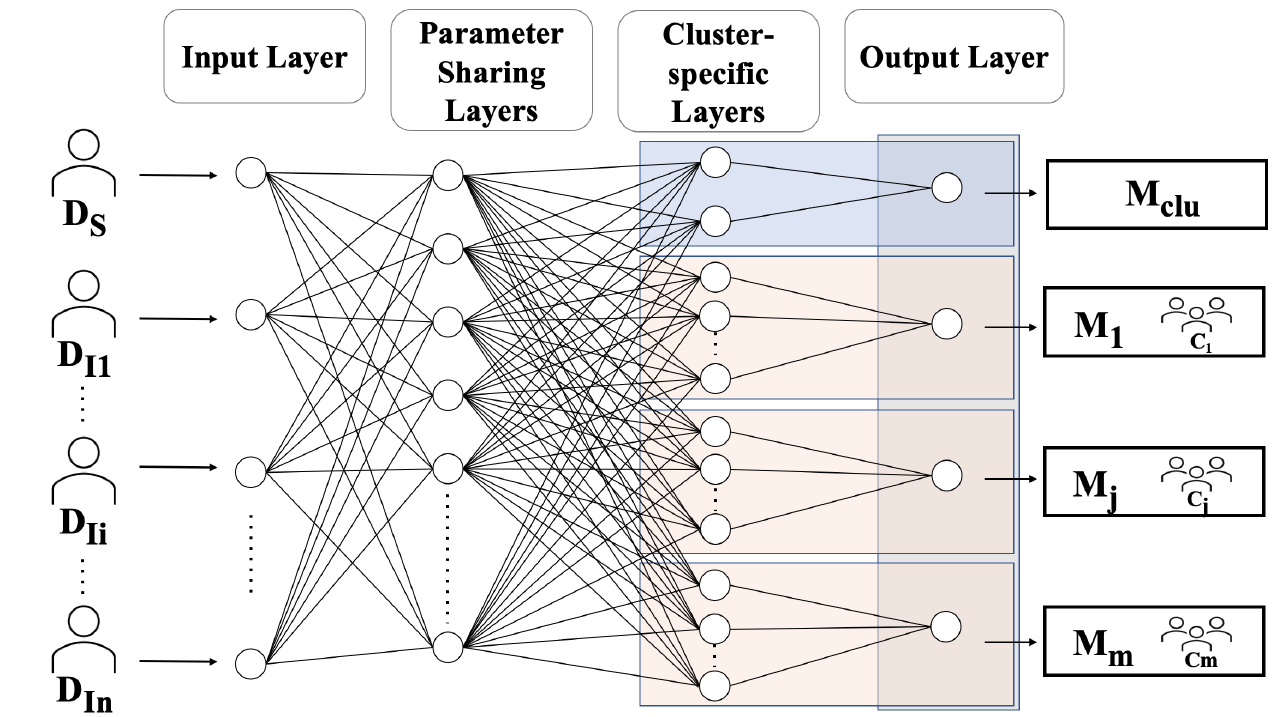}
    \caption{A demonstration of the Dual Label-Cluster-as-Tasks multitask learning approach. A model for profile clustering $M_{Clu}$  is trained on the whole dataset $D_S$ along with affective inference models ${M_1 \dots M_m}$ by optimizing a loss function across the clustering task and the affective inferences of all clusters $C_1 \dots C_m$.}
    \label{fig: Dual Label-Cluster-as-Tasks}
\end{figure}

In Dual-Label-Cluster-as-Tasks multitask Learning (cf. Figure \ref{fig: Dual Label-Cluster-as-Tasks}), given $D_S:=\left\{D_I \mid I \in S\right\}$ the set of all individual datasets, the model that is responsible for clustering and labeling $M_{Clu}$ is trained together with the cluster-specific affective inference models ${M_1, M_2, \cdots, M_i, \cdots, M_m}$ by optimizing an aggregated loss function across the clustering task and the affective inference of all user clusters $C_1, \dots, C_j, \dots ,C_m$.

\textbf{Example Approaches-- } The first technique, the multitask Multi-Kernel Learning, is proposed by \cite{jaques2016multi,taylor2017personalized} to enable information sharing through kernel weights. For each parallel optimized task, the weighted sum tasks' modalities are combined to form a single kernel function. The optimal parameter that determines the weight in the kernel is learned by maximizing a Support-Vector-Machine-like objective function with a joint constraint across all tasks. In \cite{KANDEMIR201497,jaques2016multi,taylor2017personalized}, the multitask Multi-Kernel Learning technique was developed under both the label-as-task and user-as-task scenarios to predict the next day's happiness, stress, and health through input from wearable sensors. In \cite{lopez2018multi}, the multitask Multi-Kernel Learning technique was also introduced to model cluster-as-task pain recognition. The clusters were determined by means of spectral clustering \cite{von2007tutorial}. 

The Deep Neural Networks approach of multitask learning \cite{caruana1993multitask} has been used to facilitate personalization in label-as-task and user-as-task settings.  It consists of two components: multiple shared hidden layers across tasks and unique final layers for every single task. It was proposed to deal with happiness, stress, and health detection problem in the first place \cite{jaques2016multi,taylor2017personalized} and followed by the work of \cite{saeed2018model}, who also adapted the multitask Deep Neural Networks approach in personalizing the driver stress detection model. The architecture received inputs from 16 physiological features and go through the shared layer, a subject-specific dense layer, and an output layer in order. Each training session iteratively inputs data batches from a randomly selected task and learning weights for both the shared and task-specific layers.  All \cite{jaques2016multi,taylor2017personalized,caruana1993multitask,saeed2018model} utilized hard parameter sharing in the field of multitask learning, according to \cite{Ruder17a}. It shares parameters between tasks by implementing joint hidden layers between all tasks while remaining one or more task-specific output layers. On the other side, we haven't seen any work adapting soft parameter sharing in personalized affective computing, which does not share the same hidden layers but instead allows each task to have its own parameters. The parameter sharing will be enabled through regularization, which will keep the task-wise parameters to be similar. 

The third technique is the clustering approach for multitask learning. One of the examples is the Hierarchical Bayes Model, proposed by \cite{rosenstein2005transfer}. It is made possible under the assumption that the tasks share a common Dirichlet prior. Following the work of \cite{xue2007multi, KANDEMIR201497}, a non-parameterized approach was chosen by \cite{jaques2016multi}, which clusters the users before jointly training cluster-wise logistic regression classifiers and therefore realized the personalization for Cluster-as-Task scenarios.  
In \cite{zhao2019personalized}, a similar clustering idea was adopted. By constructing a weighted vertex for each user, it clusters the users based on hypergraphs and performs personalized emotion recognition for multiple subjects from personality and physiological signals at a time. 

The approach of \cite{chithrra2022personalized} is the first work exploring personalization under the scenario of dual label-cluster as tasks. It proposes a multitask learning framework for personalizing productive engagement scores in robot-mediated learning. The approach aims to predict the engagement score and classify user groups (three learner profiles) at the same time. The joint classification of the students' profiles and the engagement score regression through multitask learning play the role of an implicit personalization framework by taking into account the profile information through weight sharing.

\textbf{Discussion.} Multitask Learning is advantageous for computational efficiency, as training one model on multiple objectives can be more efficient than separate models. It also reduces overfitting risk, as learning multiple objectives encourages more general representations to perform well across tasks, rather than overfitting to any single objective.

\subsubsection{Other} 
To the best of our knowledge, \cite{cai2018inferring} presents a unique approach not typically seen in existing categories. Recognizing a pattern where users within the same groups exhibit similar emotions in their shared images, \cite{cai2018inferring} introduced a Group-Based Factor Graph Model. This model employs a graphical structure to represent the relationships among image content features, individual demographic information, personal temporal emotion correlations, and the group's "homophily." This innovative structure facilitates effective emotion modeling on both individual and social group levels, thereby capturing the social element crucial for emotion inference.

\section{Discussion \& Open Questions}
\label{sec:discussion_openQ}

\subsection{Personalization Techniques}

To address our second research question, we conducted an analysis of the distribution of personalization techniques featured in the papers reviewed. We categorized both group-specific and user-specific models under the umbrella of "data grouping" methods. This classification is based on the commonality that both approaches involve training affective computing models focused on either individual targets or manually defined clusters. 

In our analysis, considering that some studies may employ multiple techniques from those we proposed, we accounted for each instance of a method as a separate count in our statistical analysis. However, for other aspects like yearly distribution, deep learning (DL) or machine learning (ML) analysis, each paper was counted only once, irrespective of the number of methods used within it.

\begin{figure}[!ht]
    \center
    \includegraphics[scale=0.5]{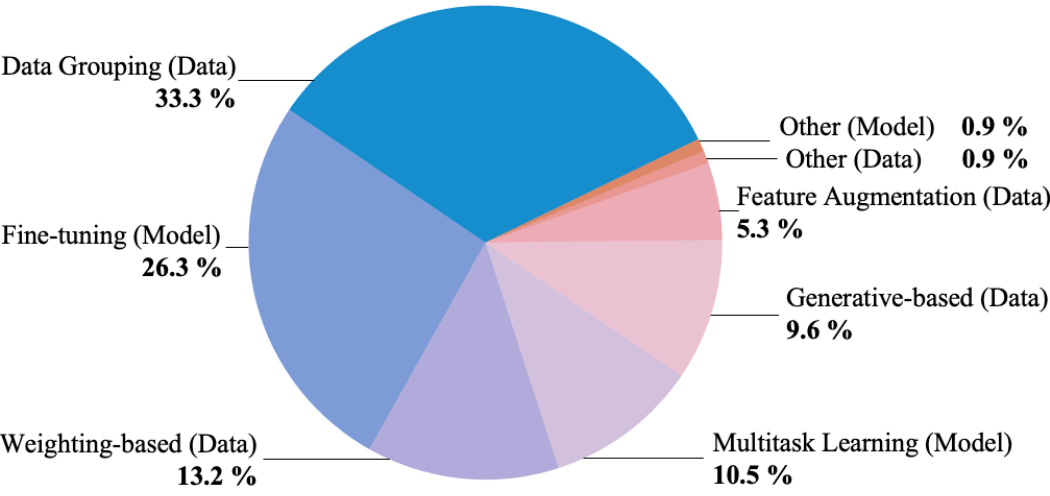}
    \caption{Distribution of the personalization techniques discussed in the surveyed papers.} 
    \label{fig:techniques_all}
\end{figure}

Figure \ref{fig:techniques_all} illustrates that the data-grouping method, which combines User-specific and Group-specific approaches, is the most commonly used technique, accounting for 33.3\% of the methods in our survey followed by fine-tuning 
utilized in 26.3\% of the studies. 
In contrast, feature augmentation is the least adopted method within our taxonomy, representing only 5.3\% of the techniques in the surveyed papers.

Data-grouping and fine-tuning approaches are common in the literature because they can be easily adjusted and customized for different datasets. They do not require the general dataset $D_G$ to be strictly representative of the target dataset $D_t$. Instead, they manually adjust the data or model for the target of interest. This resolves the problem of dissimilarity between $D_G$ and $D_t$. 
Both techniques can achieve satisfying results with fewer constraints on the underlying distributions of the dataset, and the gap can be filled easily by human efforts.

Weighting-based approaches are also widely used due to their intuitiveness and simplicity. When there is similarity between the general dataset and target dataset, or among different target datasets, weighting enables efficient reuse of pre-trained models and samples while minimizing labeling costs. However, copying models or samples requires extra storage and computing resources. This drawback helps explain why, despite their intuitiveness, weighting-based approaches have not achieved the highest popularity. Their resource demands counterbalancing their conceptual simplicity.

Multitask learning has gained some popularity too. When sufficient target-specific data is available, training models on a federated basis can help uncover relationships between individuals. Existing works using multitask learning for personalization of affective computing tasks \cite{jaques2016multi,taylor2017personalized,caruana1993multitask,saeed2018model} utilized hard parameter sharing. This approach shares parameters between tasks by implementing joint hidden layers for all tasks while maintaining task-specific output layers. However, we have not seen works adapting soft parameter sharing for personalized affective computing. Soft parameter sharing allows each task to have its own parameters. enables parameter sharing via regularization rather than complete parameter equality. This might be a potential direction for future explorations.  

Generative-based models are less commonly adapted even though they can generate representative user data. This is likely due to several challenges: First, generative models like GANs are difficult to train. Stabilizing the adversarial training process is complex and compute-intensive. Second, while generated data augments limited real samples, it may fail to fully capture nuances needed for accurate personalized models. Unrealistic synthetic data can also negatively impact classifier training. Finally, the two-step generative plus discriminative modeling pipeline creates more system complexity versus direct personalized modeling.

Finally, feature augmentation approaches account for only 5.3\% of surveyed papers. One reason is that correlations between affective states and personal attributes (e.g. gender, age) may be too weak or dataset-dependent to enable effective auxiliary feature generation. Secondly, similar to generative models, the complex two-step pipeline for augmentation and then affect prediction adds overhead that makes this approach less attractive to researchers. However, it is also worth noting that the explicit secondary features generated via augmentation can provide more model interpretability. This interpretability contrasts with opaque representations learned by end-to-end approaches. As such, when feature relationships exist, augmentation may open doors for explainable affective modeling research.

\begin{figure}[!ht]
    \center
\includegraphics[scale=0.42]{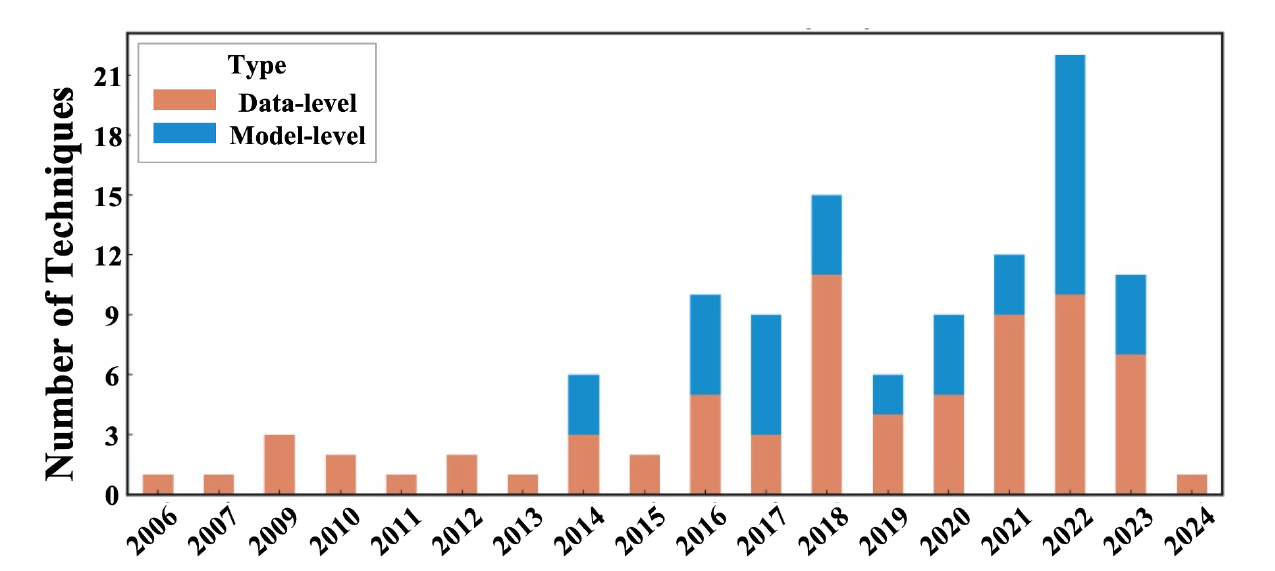}
    \caption{Distribution of existing personalization techniques in the literature by year in terms of the stage where the personalization is done: data-level vs. model-level. Model-level approaches have not been explored until 2014 and year 2022 has been the richest in terms of research about both data-level and model-level techniques.}
    \label{fig: branch}
\end{figure}

\textbf{Personalization Techniques: Data-level vs. Model-level}. In this paper, we further categorize the techniques employed in the surveyed studies into two primary categories: data-level techniques and model-level techniques. This classification is illustrated in Figure
\ref{fig: branch}. Data-level techniques typically involve applying one or more data manipulation techniques, such as categorizing, re-weighting, and augmenting the target data. This enables the training target-specific models, each of which has similar architectures. On the other hand, model-level techniques involve the training of specialized models with different architectures to capture the affective state of the group of interest without requiring further manipulation of the dataset. The distinction between these two categories lies in the approach to personalization. Data-level techniques aim to adapt the existing models to the target data, while model-level techniques construct new models for the target data. 

As shown in Figure \ref{fig:techniques_all}, 62.3\% of surveyed papers use data-level techniques while 27.7\% utilize model-level techniques. Although data-level approaches seem much more prevalent, this is mainly because of the data-grouping models that comprise 33.3\% of all surveyed methods, accounting for a major portion of data-level techniques. If we exclude standalone modeling, the model-centric strategies like fine-tuning and multitask learning comprise a greater portion than specialized data techniques like generative models or feature augmentation. The reasons that the model-centric personalization may be more frequently adopted is that data generation and augmentation add complexity without guaranteed benefits, while the model-level approaches are adaptable to new target data using standard workflows and provide flexibility while sharing and consolidating knowledge.

Figure \ref{fig: branch} helps to dig deeper in the patterns of the usage of data-level and model-level approaches as the machine learning paradigm developing. The surveyed period was set to be from 2006 to 2024. From the histogram, we can see that the number of papers published each year has generally increased over time, with some fluctuations. The most papers were published in 2022, which aligns with the increasing interest and advancements in computing models in recent years. 

Between 2006 and 2015, the domain of affective computing was predominantly focused on data-level personalization approaches, as evidenced by the literature whereas, only three works proposed model-level techniques. Commencing in 2016, a discernible shift began to emerge, with model-level approaches progressively gaining interest. This transition, marked by the increasing sophistication of computational capabilities, culminated in 2022 when the prevalence of model-level works surpassed that of data-level techniques. However, data-level personalization has maintained a consistent momentum throughout the surveyed period. Its absolute numbers indicate that it remains a popular and viable option for implementing personalization.

\textbf{Personalization Level: User-level vs. Group-level}. From the angle of the target of personalization, we have categorized the surveyed papers into two groups: user-level and group-level personalization. According to our analysis, 81.2\% of the previous works opted for personalization at the user level while 18.8\% focused on the group level. This trend can be attributed to several reasons. First, individual differences are typically greater than profile differences between groups, making it more important to tailor affective models to individuals. Second, the available datasets may not support investigations into group-level personalization due to their high diversity, rendering group differences statistically insignificant. Moreover, individual-level personalization models have been found to be more effective than group-level models in numerous studies. However, it is also important to recognize that group-level outcomes is capable of presenting valuable insights into the affective states of a larger population and that these outcomes can be used to strengthen models for user-level personalization. Therefore, future studies in personalized affective computing are suggested to prioritize exploring user-level personalization while also leveraging group-level outcomes to improve the overall effectiveness of the models. This approach would enable researchers to acquire a more comprehensive understanding of the affective states of both individuals and groups and to design more effective personalized interventions accordingly.

\textbf{Interaction Mode: HCI vs. HHI vs. HRI}. The interaction mode across all the surveyed works has also been recorded. With 80 papers focused on Human-Computer Interaction (HCI), this mode has received the most research attention by far. The predominant focus on HCI is likely due to its wider relevance and data availability compared to other modes. HCI systems leverage the abundance of user data from computers and mobile devices, providing researchers rich data to explore personalization methods. Additionally, HCI systems have very broad applicability - developing personalized affective modeling in HCI has substantial commercial, educational, and psychological motivations. Human-computer interfaces are ubiquitous, so research to personalize them has immense value. In contrast, only 13 surveyed works focused on Human-Human Interaction (HHI) and 8 on Human-Robot Interaction. While HHI and HRI are active areas of affective computing research, the challenges of collecting naturalistic human interaction data and requiring specialized robotics expertise limit broader investigation compared to HCI. Given the ubiquity of human-computer interaction, wealth of associated data, and broad motivations, HCI represents the most widely and deeply investigated interaction mode for realizing personalized affective computing.

\begin{table}[]
\centering
\caption{Distribution of the personalization context. }
\label{Context}
\begin{tabular}{@{}lll@{}}
\toprule
\textbf{Context}  & \textbf{Percentage} \\ \midrule
Emotion Understanding & 32.8\% \\
Health & 21.8\% \\
Educative  & 13.9\% \\
Social Interaction & 12.9\% \\
No Context & 5.9\% \\
Entertainment & 5.9\% \\ 
Facial Expression Understanding  & 5.9\% \\ 
Bot Service & 0.9\% \\ \bottomrule
\end{tabular}
\end{table}

\textbf{Personalization Context.} The context where personalization investigations have taken place highlights a diverse range of applications and areas of interest. As shown by Table \ref{Context}, the trend in personalization investigations is heavily skewed towards natural emotion understanding and health, with more than half of the studies investigated in this context. Significant attention is also given to education, social interaction and entertainment. Lastly, bot service, at 0.9\%, indicates an emerging interest in personalizing automated services and interactions, reflecting the increasing integration of AI and bots in daily tasks and customer service.

\subsection{Personalization Approaches and the Evolution of Machine Learning Paradigms}

Our literature review has revealed that a vast majority of personalization approaches in affective computing focus on more traditional approaches rather than deep learning methods. 
However, a pertinent question emerges concerning the evolving landscape of machine learning. Given the trajectory of the field and the notable performance gains achieved by increasingly larger models, a critical consideration arises: should our focus shift towards fine-tuning these larger models exclusively?
This question stems from a genuine skepticism regarding the longevity of the current wave of models. The pace at which machine learning is advancing prompts us to reflect on the efficacy of continually scaling models and the potential limits of this approach. While our survey illuminates the prevalence of traditional methods in personalized affective computing, it is crucial to explore whether the field is on the verge of a paradigm shift towards prioritizing fine-tuning of larger models.

As we navigate the implications of this potential shift, it becomes imperative to reevaluate existing personalization strategies. Traditional methods have demonstrated effectiveness, but in light of the unprecedented capabilities offered by larger models, a comprehensive understanding of how these methods align with or need adaptation to contemporary machine learning trends is crucial. The discussion around the longevity and scalability of current models adds a layer of complexity to the ongoing discourse on personalized affective computing, urging researchers to critically assess the path forward in the pursuit of optimal model performance and adaptability.

\begin{figure}[!ht]
    \center
\includegraphics[scale=0.55]{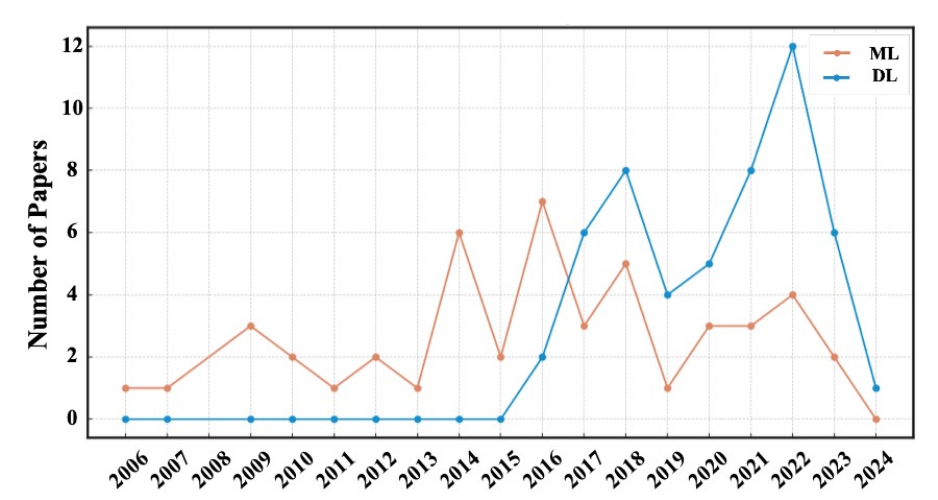}
    \caption{Trends in personalization along the evolution of Machine Learning Paradigms: machine learning (ML) vs. deep learning (DL). The use of deep learning techniques has been taking over in the last five years.}
    \label{fig: TrendsMLDL}
\end{figure}

In general, out of the total papers in the collection, 60 papers use Deep Learning, 51 papers use classical Machine Learning, and 3 papers use both DL and ML. Deep Learning (DL) is used slightly more than Machine Learning (ML) in the surveyed collection.

Figure \ref{fig: TrendsMLDL} suggests that machine learning methods have long been adapted to enable personalized computational modeling of human affect while deep learning came into play since around 2016. This is indicated by the steady rise in papers utilizing deep learning from 2016 onward. The 2018-2023 interval exhibits a 30\% drop in machine learning publications. The sharp decline of the line denoting papers using machine learning methods in 2018 may signify a shift toward deep learning techniques at that time, given deep learning's rapid ascent.

This shift could be due to several reasons. DL methods, such as neural networks, have shown to be highly effective in handling complex tasks and large datasets, which are common in affective computing. DL methods can automatically learn and extract features from raw data, reducing the need for manual feature engineering. Moreover, personalization tasks often require more complex models and network architectures to capture the nuanced differences in affect expressions. This is because affective responses can vary greatly between individuals and even within the same individual under different conditions. Therefore, models need to be able to capture these subtle variations to accurately personalize the computing performance. In the meantime, DL models can be trained end-to-end, meaning that all parts of the model are updated to improve the prediction performance.

The decline in the blue line plotting papers using deep learning may be partially attributable to a time lag - that is, some more recent publications utilizing deep learning were likely still under review and thus not yet accessible at the time of the review. 

Comparing Figure \ref{fig: branch} and Figure \ref{fig: TrendsMLDL}, we notice a parallel increase in the adoption of model-level techniques and deep learning applications within the field, around 2016 -- 2017. This can be attributed to the fact that model-level methods, such as fine-tuning and multitask learning, rely heavily on advancements in deep learning technologies. The observed trends also suggest that the future trajectory of personalization techniques in affective computing will be closely intertwined with the evolution of the machine learning paradigm.

\subsection{Personalized models and bias}
When training personalized models, user profiling with respect to personal factors is conducted. This gives rise to apprehensions about partiality and possible bias in the judgments made by personalized models towards specific social groups, such as age and gender. Studies have shown that generic models of affect can exhibit biases towards such groups if not properly addressed  \cite{XuWKG-ECCVW20,cheong2021hitchhiker}, but investigating and mitigating bias in personalized computing models has yet to be explored in literature. Furthermore, it is still an unresolved matter whether personalizing affective inference systems escalates or lessens bias and promotes fairness. Hence, further exploration is necessary to answer this question.

\subsection{The need for user-oriented datasets}
Personalized affective computing models are trained on datasets that capture the behavior, preferences, and characteristics of the target users. 
However, existing  datasets may not capture the individual variations and nuances that are crucial for personalization. 
To address this limitation, more user-oriented datasets are needed for personalized affective computing. These datasets should be designed to capture the unique attributes and preferences of individual users, such as their demographic information, behavior history, personality traits, and contextual factors. By incorporating this information, affective computing models can provide more accurate and personalized predictions.

Additionally, user-oriented datasets can also help to investigate and even mitigate bias in affective computing models. Traditional datasets can often reflect the biases and assumptions of the data creators and may not be inclusive of underrepresented groups. User-oriented datasets can be designed to be more diverse and representative of different user groups, which can help to ensure that the computing  models are fair and unbiased.

\subsection{Fairness}
Personalized models should take into consideration ethical considerations \cite{thelisson2018general} such as fairness, particularly with respect to  treating similar individuals equally. So it is important to ensure that the computing model is able to personalize equally well on data from each individual (or group if personalization is performed at the group level). This means that variances in performance metrics across individuals or groups should only be tolerable up to some extent. This constitutes an interesting avenue of research within this topic.

\subsection{Generative AI for Personalized Affective Computing}

The advent of generative AI offers a transformative approach to personalized affective computing. Unlike traditional methods requiring predefined features, generative AI tools such as Large Language Models (LLMs) can dynamically generate emotionally nuanced responses based on user interactions \cite{zhang2024affective}. This capability naturally supports personalization by adapting the content, tone and style to individual affective states, enhancing engagement in applications from mental health support to educational tutoring \cite{gu2024affective}. Research shows LLMs can improve emotional text generation by 10.9\% when guided by affective prompts \cite{li2023large}.

Furthermore, LLMs' dynamic adaptation requires real-time personalization techniques that fundamentally differ from static approaches in traditional affective computing. These challenges highlight the need for dedicated frameworks to categorize LLM-specific personalization strategies and assess their effectiveness, as current methodologies remain inadequate.

Analyzing personalization in LLMs for affective computing requires new taxonomies and methodologies due to their unique characteristics. Wang et al. (2024) identified challenges where vast, unstructured training datasets complicate isolation of person-specific features \cite{chen2024large}. In \cite{abdurahman2024perils}, the authors highlighted cultural and gender biases that may undermine personalization effectiveness across diverse user populations. The same study noted how LLMs' black-box nature obscures attribution of personalization effects to specific components, necessitating evaluation metrics beyond traditional accuracy measures.

Given these complexities, the current survey, has focused primarily on established personalization techniques such as data grouping, fine-tuning, and feature augmentation, which are better supported by existing datasets and evaluation protocols. This focus is due to the maturity of these methods and the availability of structured data, as opposed to the nascent stage of LLM-based personalization research. However, we acknowledge the transformative potential of generative AI and encourage future work to build upon this foundation, particularly in explaining and expanding the role of LLMs in personalized affective computing. 

Future research should develop frameworks tailored to LLM personalization in affective computing, addressing key challenges while leveraging their unique capabilities. Priority areas include: developing new evaluation metrics that capture LLMs' dynamic nature, investigating fine-tuning techniques using user-specific data, addressing ethical considerations including bias mitigation and privacy protection, establishing standardized benchmarks and datasets for systematic comparison, integrating multimodal capabilities to enhance emotional understanding, implementing real-time adaptation mechanisms for continuous learning, improving explainability for user trust, and conducting cross-cultural studies to ensure global applicability.

\section{Conclusion}
\label{sec:conclusion}

In this paper, we have presented a comprehensive taxonomy of state-of-the-art (SOTA) personalized affective computing methods. Our taxonomy is based on either data-level methods or model-level techniques and we considered how the different affective contexts have been leveraged with personalization frameworks. Additionally, we conducted a statistical analysis of the surveyed literature, analyzing the prevalence of different affective computing tasks, interaction modes, interaction contexts, and the level of personalization among the surveyed works. Finally, we discussed open questions and future directions in the area. Based on that, this survey shall provide a roadmap for those who are interested in exploring this direction.

\section*{Acknowledgement}
H. Salam is supported in part by the NYUAD Center for Artificial Intelligence and Robotics, funded by Tamkeen under the NYUAD Research Institute Award CG010.\\

\ifCLASSOPTIONcaptionsoff
  \newpage
\fi

\bibliographystyle{IEEEtran}
\bibliography{sample-base.bib}

 \begin{IEEEbiography}[{\includegraphics[width=1in,height=1.25in,clip,keepaspectratio]{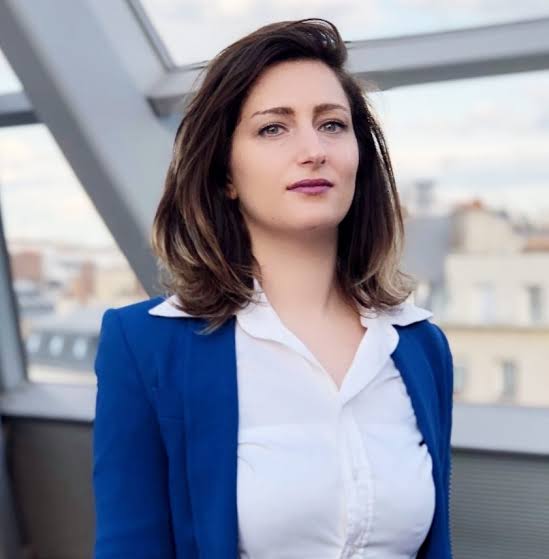}}]{Hanan Salam} is Assistant Professor in Computer Science at New York University Abu Dhabi. She is also the director of Social Machines \& Robotics Lab (SMART) \& a member of the Center of AI \& Robotics (CAIR). She is the co-founder of Women in AI, an international non-profit whose mission is to close the gender gap in the domain of Artificial Intelligence through education, research, and events. Her scientific interests include Artificial Intelligence for mental healthcare, Human-Machine Interaction, social robotics, computer vision, machine learning and affective computing. She is an advocate of technology for common good and an activist for women empowerment.
 \end{IEEEbiography}

 \begin{IEEEbiography}[{\includegraphics[width=1in,height=1.25in,clip,keepaspectratio]{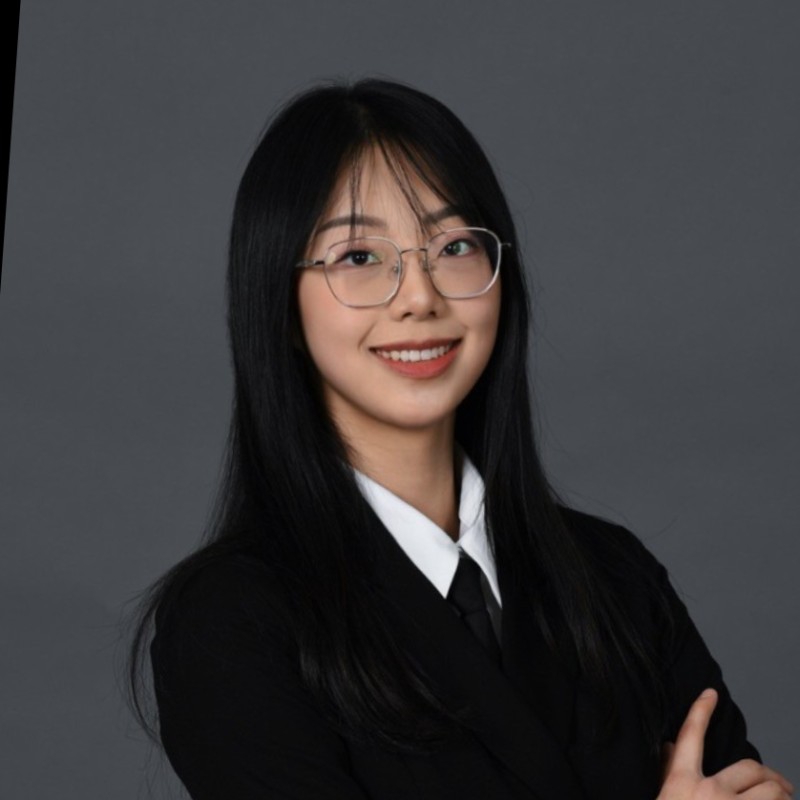}}]{Jialin Li} is a dedicated researcher specializing in machine learning, affective computing, and robotics. Holding a bachelor's degree in data science from New York University, she currently works as a research assistant in the SMART lab at NYU Abu Dhabi. Jialin's research interests include developing personalized algorithms for affective computing and integrating them into socially beneficial robots, with a focus on healthcare and education applications.
 \end{IEEEbiography}

 \begin{IEEEbiography}[{\includegraphics[width=1in,height=1.25in,clip,keepaspectratio]{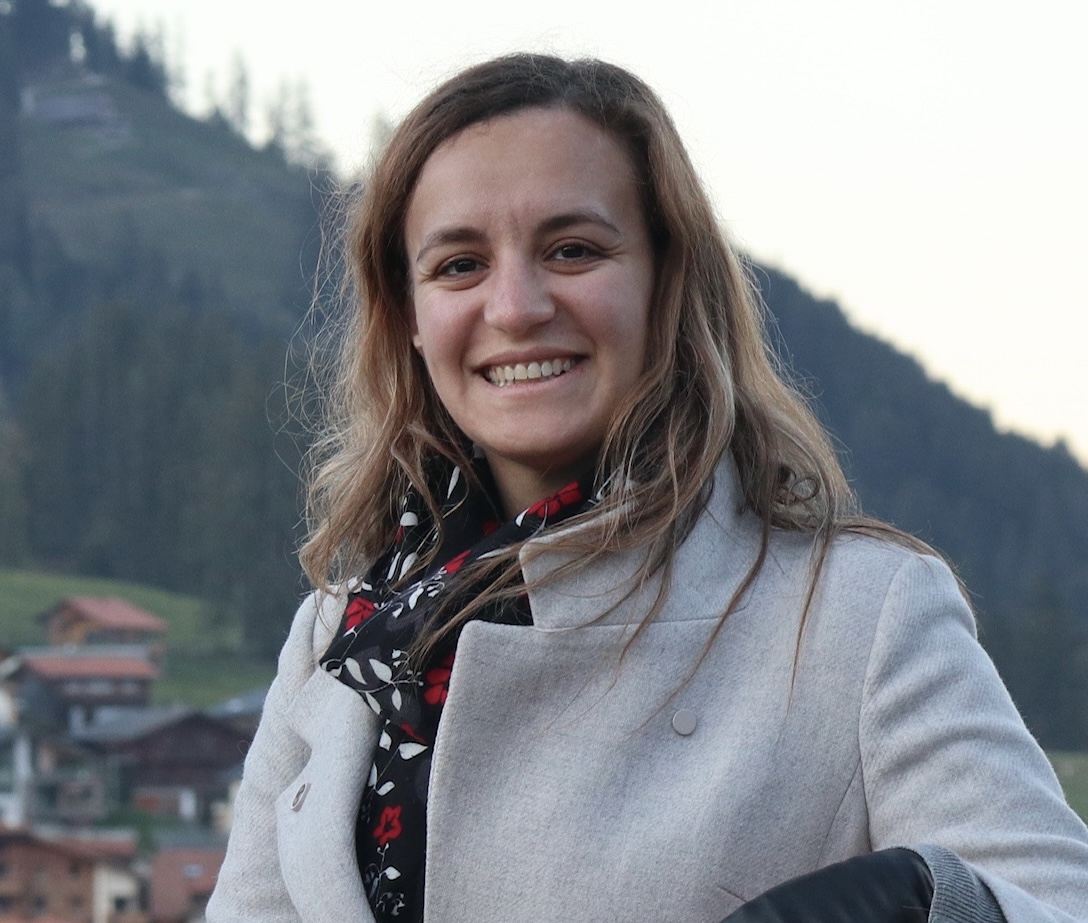}}]{Maha Elgarf} is currently a postdoctoral researcher at the SMART lab at New York University in Abu Dhabi. She holds a Ph.D. in Computer Science and Robotics from the Royal Institute of Technology (KTH) in Sweden. She previously finished her master's thesis project at the University of Augsburg in Germany. Her current research interests include social robotics, human-robot interaction, affective computing,  deep learning, conversational AI and educational human-computer interaction. 

 \end{IEEEbiography}

 \begin{IEEEbiography}[{\includegraphics[width=1in,height=1.25in,clip,keepaspectratio]{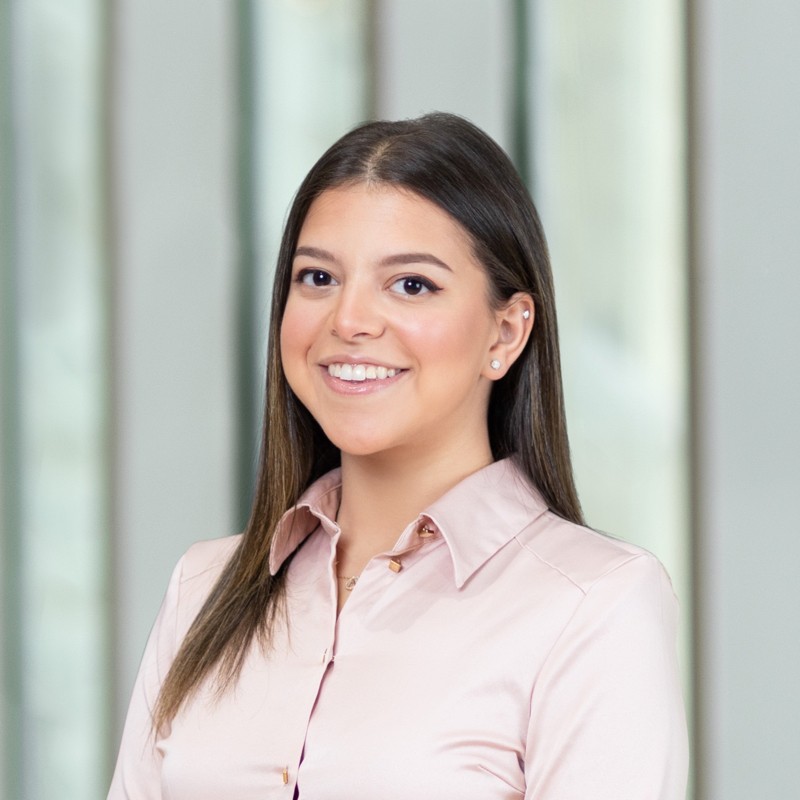}}]{Alia Waleed Mahmoud} holds a bachelor's degree in Computer Science and Interactive Media from New York University of Abu Dhabi. She worked as a research assistant at the Social Machines and RoboTics (SMART) Lab under the supervision of Professor Hanan Salam where she conducted research in personalized affective computing and student engagement detection. Here research interests are in machine learning, affective computing, and social AI.
 \end{IEEEbiography}
 
\end{document}